\documentclass{article}

\usepackage[utf8]{inputenc}
\usepackage[T1]{fontenc}
\usepackage{microtype}
\usepackage{graphicx}
\usepackage{subcaption}
\usepackage{booktabs}
\PassOptionsToPackage{hyphens}{url}
\usepackage{hyperref}
\usepackage{xurl}
\usepackage{enumitem}

\usepackage[accepted, preprint]{icml2026}

\usepackage{amsmath}
\usepackage{amssymb}
\usepackage{mathtools}
\usepackage{amsthm}
\usepackage[capitalize,noabbrev]{cleveref}

\icmltitlerunning{Lessons from External Review of a Frontier AI Safety Case}
\begin{document}

\twocolumn[
  \icmltitle{
Lessons from External Review of DeepMind's Scheming Inability Safety Case

}

  \begin{icmlauthorlist}
    \icmlauthor{Stephen Barrett}{arcadia}
    \icmlauthor{Francisco Javier Campos Zabala}{arcadia}
    \icmlauthor{Sean P. Fillingham}{arcadia}
    \icmlauthor{Umair Siddique}{arcadia}
    \icmlauthor{James Walpole}{arcadia}
    \icmlauthor{Robin Bloomfield}{arcadia,city}
    \icmlauthor{Henry Papadatos}{saferai}
  \end{icmlauthorlist}

  \icmlaffiliation{arcadia}{Arcadia Impact AI Governance Taskforce}
  \icmlaffiliation{city}{City St George’s, University of London}
  \icmlaffiliation{saferai}{SaferAI}
  
  \icmlcorrespondingauthor{Stephen Barrett}{sbarrett.work321@gmail.com}

  \icmlkeywords{AI safety, safety cases, assurance, external review, scheming}

  \vskip 0.3in
]

\printAffiliationsAndNotice{}

\begin{abstract}
Safety cases for frontier AI systems should provide a convincing argument, supported by evidence, that the risk of harm is within an acceptable bound. When developers author their own safety cases, confirmation bias and conflicted incentives can affect the quality of argument. External review can help to address this. In this paper, we apply the Assurance 2.0 framework to perform an external review of Google DeepMind's public scheming inability safety case. We surface substantive new concerns that materially affect the scope of the safety case and its applicability for decision-making. Based on this experience, we provide concrete recommendations for how external review should be conducted and what information AI developers should provide to support it.
\end{abstract}


\section{Introduction}
\label{sec:intro}
AI-based systems are already capable of causing significant harm \cite{bengio2026, stellingEvaluatingAICompanies2026}, yet safety practices are still immature compared to other safety-critical industries. To address this gap, there is interest among AI developers and AISIs in the use of safety cases \cite{SafetyCasesAISI,FrontierSafetyFramework}. Safety cases can have a number of purposes: supporting internal developer decision-making, enabling independent review, and communicating deployment risk to users and regulators. Independent external review tests whether the argument holds and gives decision-makers a basis for trust beyond the developer's self-assessment. 

Our work had two objectives: first, to investigate the value of providing an external review of a frontier AI safety case; and second, to provide recommendations on how to perform effective external review. We apply a state-of-the-art safety assurance methodology, Assurance 2.0 \cite{bloomfieldAssurance20Manifesto2021}, and illustrate how this methodology can be successfully applied by external reviewers to an example frontier AI safety case, specifically the scheming inability safety case provided by Google DeepMind (GDM) \cite{phuongEvaluatingFrontierModels2025}. We found that the structured Assurance 2.0 based methodology enabled us to successfully surface important concerns with the GDM safety case.

The closest related work that we are aware of is METR's work reviewing Anthropic's risk reports \cite{METRreview2025, ReviewAnthropicSabotage2026}. Our work differs in that we had limited engagement with the AI developer and we are providing review process recommendations. We expect our work to be of interest to policymakers and policy advisors, both within AI developers and externally, such as government regulators, AISIs, and third-party organisations undertaking frontier AI safety case review.

External scrutiny of safety cases is essential to their function as accountability instruments. This analysis was enabled by GDM's public disclosure, a practice we believe should become standard across frontier AI developers.

\section{Methodology and background}
\label{sec:methods}
Safety cases are structured arguments, supported by evidence that a system is safe enough to deploy in a given operational context \cite{buhlSafetyCasesFrontier2024}.  We use the Assurance 2.0 framework and the Declare guidance within the Claims Argument Evidence (CAE) framework \cite{DeclareGuidanceCAE}.  Assurance 2.0 provides a rigorous and systematic approach to developing, presenting, and examining assurance cases to support indefeasible confidence in safety or other critical properties \cite{bloomfieldAssurance20Manifesto2021, bloomfieldAssurance20Nutshell}.  Such indefeasible confidence ensures that no credible doubts remain that could change the conclusion.

Our review team consisted of six individuals, each dedicating approximately one day a week for three months ($\sim4$ person months total, $\sim1.5$ FTE for 3 months). Engagement with GDM was limited, consisting of some clarification of the intent behind the top-level claim prior to the project start and providing an opportunity to comment on a draft of the paper near the project end. Our review comprised the following steps:

\begin{itemize}
  \item Foundation building, including background reading on safety cases, CAE methodology, and the Assurance 2.0 framework.
  \item Initial unstructured review of the safety case, with each team member independently reading the paper and identifying concerns (see Appendix~\ref{app:initial_review_findings}).
  \item Initial structured analysis of the case's scope, stated purpose, and supported decision, together with CAE-compliance and logical-validity checks.
  \item Construction of a model representation of the assured system.
  \item Risk pathway development through group brainstorming and elaboration of four selected pathways.
  \item Parallel work streams to weigh evidence and challenge the underlying theory, including searches for counterclaims, literature review, and confirmation-theoretic assessment of the stealth and situational-awareness evidence.
  \item Development of counter-cases, starting from a brainstorm of plausible alternatives and then refining the three most promising arguments using CAE structure.
\end{itemize}
See Appendix~\ref{app:review_process} for a more thorough discussion of team formation and review process.

\section{Overview of the GDM Safety Case}
\label{sec:GDM}
Our process recommendations are exemplified through application to GDM's scheming inability safety case \cite{phuongEvaluatingFrontierModels2025}, which seeks to establish that Gemini 2.5 Pro lacks the ability to cause severe harm through covertly pursuing objectives misaligned with developer intentions.  In one branch of their argument GDM make the claim that to cause severe harm a misaligned AI needs both top human-level stealth ability AND situational awareness, citing as evidence findings in the literature and their own threat modelling.  Whilst in the second branch of the argument they claim that the AI does not have top human-level stealth ability OR sufficient situational awareness, citing as evidence, evaluations on proxy tasks.

The paper makes a clear top-level inability claim that avoids the harder problem of proving alignment directly. The structured argument is clearly presented, the evaluations are thoughtful, and the decision to open-source them is commendable. Compared with most public frontier-model deployment safety justifications, it is unusually explicit about its argument structure, its dependence on empirical evaluations, and some of its own possible defeaters. It provided us with an excellent concrete argument to inspect.

\subsection{Structured Review of Safety Case Scope and Context}
\label{subsec:scope_and_context}
To understand the scope of the safety case, we reviewed the top-level claim against the expectations given in \cite{bloomfieldCAEMiniguide3} that were developed based on experience from other safety-critical industries.

\subsubsection{Safety Case Purpose and Decision Support}

\textit{Process recommendation: Reviewers should seek an explicit description of the purpose of a safety case and the decisions it is intended to support.}

In addition to the GDM safety case, we studied the GDM Frontier Safety Framework (FSF) \cite{FrontierSafetyFramework}  and the Gemini 2.5 Pro model card \cite{Gemini25Pro} (see Appendix~\ref{app:safety_case_scope}) . Based on the available evidence, it was unclear whether the safety case was used in the deployment decision for Gemini 2.5 or whether it was intended only to be an exploratory research artifact. 

\textit{Process recommendation: Reviewers need to determine the portion of the risk universe the safety case is intended to cover, and ideally understand how the case integrates with other safety cases, so as to build a firm understanding of safety case scope.}

Despite a very broad top-level claim, the actual intent of the GDM safety case appears to be to cover severe harm arising only from misalignment risk. The FSF indicates that Critical Capability Levels (CCLs) are evaluated for three risk categories: misuse, ML R\&D, and misalignment. However, these categories do not cover the complete risk universe. For example, risks caused by combinations of those categories, such as combined misuse and misalignment, are not covered. Similarly, in other work \cite{shahApproachTechnicalAGI2025}, GDM highlight other categories of risk including risk from mistakes and multi-agent structural risks that are not covered by their FSF. From \cite{phuongEvaluatingFrontierModels2025} it was unclear to us, for example, whether human misuse (combined with AI scheming) or harm arising from mistakes were intended to be in or out of scope. 

\subsubsection{Definition of Assured System Property and Justification of Acceptability Criteria}

\textit{Process recommendation: Reviewers should obtain a detailed definition of, and justification for, the claimed safety property.}

The system property in GDM's top-level claim that is being assured is safety, specifically the avoidance of severe harm. However, \cite{phuongEvaluatingFrontierModels2025} does not define what ``severe harm'' is, and does not justify why assurance is not also provided for lower severity level harms. In \cite{shahApproachTechnicalAGI2025}, GDM defined severe harm as ``incidents consequential enough to significantly harm humanity'' and state that ``the question of whether a given harm is severe isn’t a matter for Google DeepMind to decide; instead it is the purview of society``.  However, our view is that in the absence of society-wide consensus on severe harm, the onus is on GDM to make it clear what scale of harm is in scope of the safety case. It is not possible to conduct an external review unless the system property being assured is fully defined. For the purposes of our review, we therefore adopted our own definition of severe harm based on the literature, defining severe harm as losses above \$1 billion, 100--1000 deaths, or thousands of casualties arising from a single incident (see rationale in Appendix~\ref{app:safety_case_scope}). 

\subsubsection{Description of the Assured System}

System models are abstract representations of the assured system or deployed environment that support reasoning about safety \cite{bloomfieldCAEMiniGuide7}.

\textit{Process recommendation: Reviewers should be provided with a model of the assured system that was explicitly or implicitly assumed in the construction of the safety case.}

 In the GDM case, a system model would be expected to indicate for example whether the assured 'AI model' corresponds to just the LLM, or whether it also is supposed to include scaffolds, long-term memory support, tool access, and guardrails etc. If GDM have a system model it was not provided in the paper. 

\subsubsection{Description of the System Environment}

\textit{Process recommendation: Reviewers should obtain models of the system environment into which the assured system will be deployed, and which will have been either explicitly or implicitly assumed during safety case development. One type of such system model is the functional control structure discussed by \citet{levesonSTPAhandbook}}

GDM states that the safety case is applicable to ``internal deployment.'' Concerns are raised about threats from use of the AI model in AI R\&D and safety work, so we infer that the model may be used by GDM employees, but comprehensive details are not provided about which teams will use the model, for what purposes, or what corresponding oversight mechanisms apply. It is not clarified whether ``internal deployment'' also covers Google more broadly, with its approximately 185{,}000 employees and responsibility for world-scale technologies such as Android, Chrome, Search, Gmail, Maps, and Google Cloud.  This is essential to allow any assessment of whether the confidence in the case is commensurate with the risks posed: mitigating severe harms in conventional systems would require specialised development practices and multiple defences in depth.

\subsubsection{Duration of Safety Case Validity}

\textit{Process recommendation: Reviewers should obtain details of, and justification for the duration of safety case validity.}

The Gemini 2.5 Pro model card states that alert thresholds are such that if an AI does not reach the alert threshold for a CCL, then models developed before the next regular testing interval are assumed not to reach that CCL either. However, it can be noted that the characteristic of situational awareness, and some of the characteristics underlying stealth and deception are actually desirable properties for AI product utility (Appendix~\ref{app:safety_case_scope}). A fundamental concern with the GDM safety case is that the AI developer will be actively working to introduce many of the underlying capabilities that could undermine the safety case.  The safety case should therefore include a sub-argument that explains what observed behavior would result in it becoming invalid.

\subsection{Supporting System Safety Engineering Artifacts}
\label{subsec:safety_eng_artifacts}
\textit{Process recommendation: Reviewers should obtain all critical system-safety-engineering artifacts that support the safety case, at minimum those concerned with identifying hazards, quantifying risks, and deriving associated safety requirements.} 

In conventional safety-critical systems engineering there are a number of steps that are performed in order to determine the set of system safety requirements. The safety case can then be used to provide assurance of successful identification of all requirements and their implementation. A typical set of steps might be expected to include risk assessment, determination of risk mitigation targets, classification of criticality and associated required assurance rigour, fault identification and quantitative residual failure analysis (see Appendix~\ref{app:system_engineering_artifacts} for more details). In the case of the GDM safety case, if the associated artifacts exist, we did not have access to them, since they were not published along with GDM's paper.

\subsection{Assessment of Risk Pathways and Criticality}
\label{subsec:riskpathways_criticality}
\subsubsection{Risk Pathways}
\label{subsubsec:riskpathways}

\textit{Process recommendation: Reviewers should perform their own risk pathway analysis and contrast this with the risk pathway analysis supplied by the AI developer, and assess impacts where differences exist.}

A robust safety case requires not just an argument that a system is safe, but an understanding of how it could cause harm. This section presents our analysis of risk pathways, the routes through which an internally deployed frontier AI model could cause severe harm, and explains why this analysis is essential to evaluating GDM's safety case.

We analyse risk pathways for three reasons. First, to calibrate the rigour we demand of the safety case: if plausible pathways to severe harm exist, we require greater confidence in the top-level claim. Second, to test the completeness of GDM's threat model, which rests on the theory that severe harm requires the AI to have developed both situational awareness and stealth. If we can identify credible pathways that do not require these capabilities, the theoretical foundation of the safety case is weakened. Third, to systematically enumerate the safety-relevant systems whose failure could lead to harm, enabling us to assess whether GDM has identified and mitigated all relevant hazards.

Our analysis identifies two classes of pathway. Precursor pathways describe how a currently deployed model could weaken GDM's safety infrastructure, whether through intentional undermining by a misaligned model or through organisational drift caused by over-reliance on AI, creating vulnerabilities that a successor model later exploits. Harm-materialisation pathways describe how a model directly causes severe harm, either through autonomous misaligned action or through exploitation of an unsafe operating envelope created by human error. Any harm-materialisation pathway may be preceded by a precursor pathway, which amplifies its probability and severity by degrading the defences that would otherwise intervene.

The potential consequences are severe and include disruption of Google services at scale, cyberattack on critical national infrastructure, manipulation of high-leverage communications, disabling of deployed AI safety mechanisms, facilitation of bioweapon production, and financial-system disruption. Mundane unreliability rather than misalignment could cause severe harm.

\subsubsection{Criticality}
\label{subsubsec:criticality}

\textit{Process recommendation: Reviewers should make their own assessment of safety case criticality and associated level of required safety-engineering rigour.  This should be compared with the AI developer's provided criticality analysis and corresponding applied level of rigour.}

A fundamental question when reviewing any safety case is whether the level of rigour applied is proportional to the criticality of the hazard being addressed. A system whose failure could cause a fatality demands far more stringent evidence, mitigations, validation, and verification than one whose failure causes only minor inconvenience. GDM does not provide this analysis. 

In the absence of this information from GDM, we made our own best-effort assessment, taking into account our findings from the risk-pathway analysis. Criticality is determined by severity of harm, exposure to harm, and controllability. GDM states that its concern is with severe harm; in the ASIL \cite{ISO2626292018} framework familiar to automotive engineers, this would correspond to a high severity classification. Exposure depends on the interpretation of internal deployment and could be high if the system is deployed across Google, where the AI may be used in high-leverage systems such as Search, Android, and Chrome. Controllability is uncertain, and oversight might be circumvented, as the paper itself recognises. This combination points toward high criticality. See Appendix~\ref{app:criticality} for more details. 

\subsection{Structured Review of GDM Safety Case Argument}
\label{subsec:structured_review}
\subsubsection{CAE Compliance and Logical Validity}

The key question in this section is whether the argument as presented meets ordinary expectations for CAE structure and whether there is sufficient information to support assessment of logical validity, confidence and residual doubt.

\textit{Process recommendation: Reviewers should assess safety cases for CAE-compliance and logical validity.}

Our analysis identified many non-compliances with CAE including, incomplete top claim, use of a disjunctive argument, inadequate definition of terms, missing side warrants, unresolved defeaters and lack of correct evidence incorporation (see Appendix~\ref{app:cae_compliance}).

\textit{Process recommendation: Reviewers should work through the entire safety case argument to search for and enumerate defeaters.}

In a second review step, two team members worked to consider the logical validity of each and every element of GDM's argument. Each team member considered a separate branch of the safety case and then critiqued each other's work. Findings (see Appendix~\ref{app:logical_validity}) were captured in the form of a list of defeaters, which were used to annotate GDM's safety argument. This form of annotation using defeaters can also be used for flagging doubts other than those just arising from logical validity.

\subsubsection{Assessment of Theory}
\label{subsubsec:assessment_theory}
Assurance 2.0 includes the concept of theories. Theories are self-contained technical descriptions and assurance arguments for specific assurance methods. Much of a safety case could in principle be synthesized from a library of such parameterised theories \cite{bloomfieldAssurance20Nutshell}.  

\textit{Process recommendation: Reviewers should evaluate the appropriateness and correct implementation associated with any theories used to justify the argument steps in  the safety case.}  

To review the theories within a safety case an external reviewer would need to check that the context in which the theory is applied is valid and that any criteria necessary for the theory to be successfully implemented have been satisfied. Since the GDM case did not cite any external theories we did not undertake this step.  

However, GDM's safety case does incorporate a number of internal and implicit theories. One of these theories (we labelled 'Theory 1'), is the theory that, to cause harm in deployment Y a misaligned AI needs both top human-level stealth ability AND basic situational awareness. We investigated this theory as now described through both a search for counter-examples and by assessment of the literature that was cited as supporting evidence.

\paragraph{Counter-examples to Theory 1}
We first searched for direct rebutting counter-examples to Theory 1.  We found that the claim that severe harm requires top human-level stealth ability is not established. Historical counter-examples show that severe harm can arise from a single accepted change under ordinary oversight. Two especially clear cases are the CrowdStrike outage of July 19, 2024, where a faulty content-configuration update caused global disruption through widespread Windows crashes affecting 8.5 million systems and causing an estimated \$2 billion or more in losses in the UK alone \cite{wellsCrowdStrikeITOutage2024}. Likewise the Knight Capital incident of August 1, 2012, where an erroneous code deployment caused more than four million orders to be sent into the market within 45 minutes, producing losses of more than \$460 million (\cite{SECgovSECCharges}). In neither case was anything like top-human covert behaviour required; what mattered was access to a privileged operational pathway and the failure of ordinary controls. This directly pressures the child claim that humans under default oversight cannot cause severe harm, and therefore weakens the inference that a misaligned AI would need top human-level stealth to do so.

\paragraph{Assessment of evidence provided in support of Theory 1}

\textit{Situational Awareness.}
Section 2.2 of the GDM paper states that to cause severe harm, misaligned AI would need at least basic situational awareness. The paper labels the supporting evidence as expert consensus and cites two sources as addressing the necessity of situational awareness for severe harm: \citet{cotraSpecificCountermeasuresEasiest2022} and \citet{ngoAlignmentProblemDeep2024}.   GDM also cite a second group of sources,  not to support the claim that some situational awareness is necessary, but to frame disagreement about the level and type of situational awareness required.  Overall, our analysis (Appendix~\ref{app:support_of_theory}) finds these sources support at most the weaker proposition that situational awareness is a graded and safety-relevant capability, not the stronger claim that ``basic'' situational awareness is the right minimum threshold for severe harm.

\textit{Humans Subject to Default Oversight Cannot Cause Severe Harm.}
GDM claim that ``Humans subject to default oversight mechanisms cannot cause severe harm'' and labels its supporting evidence as ``threat modelling and examples.'' On inspection, however (see Appendix~\ref{app:support_of_theory}), the paper does not provide formal threat modelling, worked examples, case studies, or empirical evidence for this premise. The closest it comes is a short paragraph asserting that oversight of human employees has historically been sufficient to prevent all but the most skilled humans from sabotaging or taking over the company, and inferring from this that an AI would need top human-level stealth if subject to comparable oversight. As presented, this is better understood as an intuitive analogy than as a substantively evidenced claim.  

\subsubsection{Weighing of evidence}
\label{subsubsec:weighing_evidence}

\textit{Process recommendation: Weighing evidence using confirmation theory can give the reviewer a deeper understanding of the merits of the evidence. The two team members who undertook this activity found the process itself valuable in forcing a structured examination of the evidence from different angles, but the outputs were not directly useful. If time permits, it is worth attempting.}

The safety case depends heavily on behavioural evaluations. The relevant question is not simply whether a model passed or failed particular tasks, but how strongly those outcomes support the inability claims used in the safety case. Our analysis is confirmation-theoretic in structure \cite{bloomfieldAssessingConfidenceAssurance2024}: for each local claim $C$ and body of evidence $E$, we ask whether $E$ would be more expected if $C$ were true than if it were false. Numerical confirmation measures such as likelihood ratios or log-likelihood-ratio scores can formalise this comparison, but we do not report point estimates in the main text. This is because the required likelihood assignments are highly subjective, the evidence items are overlapping and non-independent, and the main disputes concern construct validity and transfer to deployment rather than precise aggregation.

We therefore report our judgement qualitatively using five categories: strong confirmation, moderate confirmation, weak confirmation, mixed or neutral, and disconfirmation. These categories indicate comparative evidential force, not posterior probabilities that a claim is true.

Across both the stealth and situational-awareness branches, we assess the evidence (see Appendix~\ref{app:weigh_evidence}) under the same three dimensions .  Firstly, proxy validity, which concerns whether the evaluation suite reasonably operationalises the target capability.  Secondly, benchmark performance, which concerns whether the observed results robustly show poor performance on that suite.  Thirdly, transfer, which concerns how far those results support the stronger claim of deployment-relevant incapability.

We found that the stealth evidence provides moderate confirmation for the claim that Gemini 2.5 Pro does not robustly exhibit top human-level stealth on the tasks GDM designed. It provides only weak confirmation for the stronger claim that the model lacks deployment-relevant top human-level stealth.  Whilst, the situational-awareness evidence provides moderate confirmation for the narrow claim that current models do not reliably display the specific situationally aware, constraint-lifting behaviours measured by this suite. It provides only weak confirmation for the stronger claim that they lack deployment-relevant basic situational awareness.

\subsubsection{Residual Doubt}
\label{subsubsec:residual_doubt}

\textit{Process recommendation: Where the system is safety-critical, the reviewer should check that the safety case is complete and that any remaining defeaters should represent residual doubts for which a risk assessment has been made, communicated, and accepted. To mitigate bias, reviewers may wish to make their own determination of residual risk before reviewing the developer's assessment.}

GDM states that the defeaters they themselves identified are important, but that nevertheless their safety case will hold in the near future. We assume therefore that they regard their defeaters as representing residual risks they have deemed acceptable. However, they provide neither quantification of those risks nor an explanation of why they are acceptable.  If a safety case has unmitigated defeaters then the residual risks cannot be accepted. 

\subsubsection{Confidence}
\label{subsubsec:confidence}

\textit{Process recommendation: Reviewers of safety cases should request the AI developer to provide an assessment of confidence in the top-level claim, or at least their assessment of which parts of the safety case they have judged to be weakest. Reviewers would ideally form their own judgement of which parts of the safety case are weakest and most confidence impacting prior to reviewing the developer's assessment.}  

As noted earlier the GDM case did not provide any analysis of the criticality of the claims and the confidence needed in them. Confidence assessment normally has most value when the case has been shown to be logically sound and valid and defeaters have been mitigated. As this was not the case we did not explore this further.  

\subsection{Counter-Cases}
\label{subsubsec:countercase}

\textit{Process recommendation: Counter-cases can be used to both identify new defeaters, doubts and concerns and to form a more robust argument, linked through to evidence for why a safety case may not be adequate to justify deployment. We expect that such counter-cases could represent another good vehicle for dialectical engagement with the AI developer.}

Following the suggestion of \cite{bloomfieldCAEMiniGuide7}, we developed 3 counter-cases that challenge load-bearing parts of the GDM argument (Appendix~\ref{app:counter_cases}):

\subsubsection{Counter-Case 1: Unjustified Use of Evaluations}

This counter-case argues that GDM's claim that the evaluation results can adequately assure absence of basic situational awareness and top human-level stealth is unjustified. The central problem is not a single weak evaluation, but a combination of factors. GDM provides no proof of absence of sandbagging (\cite{naikAgentMisalignmentMeasuringPropensity2025}), the situational-awareness suite is narrow and the stealth evaluations are less independent than the raw count suggests.  There is also a lack of clarity on the boundaries of  the system that is being evaluated, and how that system might be instantiated in deployment considering potential presence of affordances such as tooling and accumulated context.  This critique also exposes a deeper epistemic gap. The evaluations observe behaviour in specific tasks, but the safety claim concerns the absence of dangerous internal capabilities. More fundamentally, based on experience in safety critical industries, behavioural evaluations of black box models cannot on their own provide the level of assurance that a high-criticality deployment decision would require \cite{bloomfieldAssuranceAISystems2025} . 

\subsubsection{Counter-Case 2: Unjustified Reliance on Human Oversight}

This counter-case builds on and adds to a number of related defeaters that GDM themselves identified. It highlights that the paper does not adequately specify what the oversight regime actually consists of in operational terms. It finds that the analogy between AI systems and human employees is weak in several safety-relevant respects. AI systems can act faster, at greater scale, with more common-mode correlation, and with output volumes that may exceed realistic human review capacity. In addition, human oversight is itself vulnerable to predictable failure modes such as automation bias, alert fatigue, time pressure, and manipulation by model outputs. 

\subsubsection{Counter-Case 3: Unjustified Reliance on Product Assurance Only}

The central claim of this counter-case is that deployment decisions for AI systems capable of causing severe harm cannot be justified through product assurance alone. While capability evaluations provide valuable point-in-time evidence about a model's behaviour, they are fundamentally insufficient without accompanying process and organisational assurance. This reflects established practice across safety-critical industries such as aviation, automotive, nuclear, and medical devices, where product testing is never considered sufficient on its own (e.g. \citet{ISO2626272018} and \citet{Annex19Safety}). The argument identifies four independent pathways through which a product-only safety case can fail, any one of which is sufficient to undermine confidence in deployment safety.

\section{Summary of Review Findings}
\label{sec:summary_of_review}

We applied the standards of external review appropriate where the safety case is used to support a deployment decision rather than a research use. As external reviewers we came with the mindset of trying to show that the assured system is not safe to deploy as suggested by Leveson \cite{levesonEngineeringSaferWorld2012}. This is a powerful technique for countering the confirmation bias of the creators of the safety case. 

Our assessment found that the combination of, a) the concern with high severity of harm, b) the broad exposure to hazardous operational situations, and c) uncertain controllability all pointed clearly toward a high-criticality problem requiring rigorous, safety engineering and well-founded argumentation.

Against this criticality, we identified gaps between what would be expected in established safety-engineering practice and what the safety case delivers, examples include:

\begin{itemize}[leftmargin=*]
  \item Lack of clarity on purpose of the safety case and decisions supported
  \item Under-specified top-level claim
  \item Lack of adequately specified abstract representations (systems models) of both the assured AI system and its deployed socio-technical environment with associated oversight mechanisms.
  \item Limitations in evidenced hazard and risk assessment and derivation of safety requirements
  \item Lack of CAE compliance
  \item Presence of defeaters indicating incompleteness of the case and meaning that neither residual risks or confidence could be assessed   
  \item Cited evidence not sufficiently supportive of key theories
  \item Evaluation related evidence providing weak confirmation of claims
  \item Absence of a proof for lack of sandbagging on evaluations.
\end{itemize}

Our work on counter-examples and counter-cases surfaced a number of insights, for example:
\begin{itemize}[leftmargin=*]
  \item The 2024 CrowdStrike incident in which what mattered was
access to a privileged operational pathway and the failure
of ordinary controls, directly pressuring the claim
that humans under default oversight cannot cause severe
harm
  \item Lack of process and organisational safety assurance, needed to complement the provided product safety assurance
  \item The evaluations observe behaviour in specific tasks, but the safety claim concerns the absence of dangerous internal capabilities.
  \item  Concerns that behavioural evaluations of black box models cannot on their own provide the level of assurance that a high-criticality deployment decision would require
  \item Weaknesses in assumptions about efficacy of human oversight
\end{itemize}

\section{Assessment of the value of the review}
\label{sec:assessment_of_value}

The value of external review lies in broadening the range of challenges applied to a safety case, revealing blind spots that internal teams may normalise, and helping determine whether a safety case is genuinely decision-ready. 
Our work demonstrates both the value and the feasibility of such review. 
Even a limited independent assessment that applies a systematic methodology such as Assurance 2.0 can surface issues that materially affect the scope, meaning, and defensibility of the safety conclusion.

External review provides scrutiny of both the composition of the assured AI system and the broader socio-technical context in which deployment occurs. 
This shift can expose issues in theory and evidence alike, with implications for both decision making and system design. 
On the decision side, review supports a clearer conclusion about what a safety case does and does not justify. 
On the design side, it can point toward concrete improvements such as tighter claim wording, more deployment-realistic evaluations, explicit system modelling, clearer treatment of sandbagging and other unresolved evaluation uncertainties, more diverse evidence, and a need for process assurance. 

\section{Additional Process Recommendations}
\label{sec:additional_process_recommendations}

A review needs structure and expertise to effectively probe and challenge the case. Detailed recommendations about review structure and each activity we undertook have already been supplied throughout the paper. Hence, in this section we provide some higher-level observations and aspects not covered elsewhere (a collected set of recommendations for AI developers is available in Appendix~\ref{app:recommendations_ai_developers}.

\textbf{Dialectic argumentation and choice of steps to undertake}. The set of review steps that should be undertaken, and their resourcing will depend on the specifics of the safety case being reviewed, the information provided by the AI developer and the nature of the engagement with the AI developer. In our case we only provided a first response to a published safety case.  Other review projects might provide for a multi-step dialectic back and forth between AI developer and reviewer. In our case due to the lack of a tight closed loop with GDM, we spent effort augmenting what had been provided, for example as regards hazard and risk pathway identification, criticality assessment and system modelling. If instead there had been a tight loop with GDM, one option would have been to  quickly reject the safety case on the grounds that were discovered very early on and have requested other information that we considered missing, such as the full risk pathway analysis, and system modelling information before spending additional time in review.   

\textbf{Building understanding within the review team.} Whilst time-consuming, there are benefits to external reviewers performing some analysis themselves for certain steps such as risk pathway modelling, criticality assessment and weighing of evidence. One reason is that it enables the reviewer to  develop a deeper understanding, another is that it is important to structure the review to avoid group-think and availability bias.   

\textbf{Avoiding group-think.} When external reviewers first perform their own assessment of certain topics such as risk pathway identification, criticality assessment and weighing of evidence they avoid being biased by what the AI developer, or even others within the review team have to say on such matters. We feel that an approach of individual assessment followed by group discussion can achieve greater diversity of thought, whilst also benefiting from collectively generated insights. 

\textbf{Safety case authorship.} We recommend that safety cases are developed within a no-blame safety culture where challenge and review is accepted as healthy. Organisations should take responsibility for the cases and should act as author. 

\textbf{Open review.} We believe there is substantial value in enabling some form of open review, even where parts of the evidence or system description must remain confidential. Open publication allows broader scrutiny, helps build better review norms across the field and ultimately should lead to safer systems. 

\textbf{Controlled disclosure.} Some information may legitimately need to be restricted for security or competitive proprietary reasons. Where this is the case, developers should state that additional evidence exists, explain the role it plays in the argument, and, where possible, make it available under controlled access to trusted external reviewers. The level of access required will depend on criticality of the deployment decision.  Going forward as both capabilities and criticality increase, improved access will become increasingly important.

\textbf{Characteristics of the review team.} Our team of 6 had domain expertise in system safety assurance including assurance of AI systems, AI safety governance, technical aspects of machine learning and AI evaluations. We all have strong scientific and technical backgrounds. For some, this was their first work on safety cases. We would anticipate that most successful review teams would need a similar set of skills and experience. Our team benefited from diversity of experience, perspectives, and imagination.  

\textbf{Timing and duration of review.} Our review required 4 person-months for the completion of one round of response in a dialectic review. AI developers are clearly working under intense commercial pressure to release AI products and no doubt would seek a quick cycle of review. Whether such a quicker review cycle is possible, how many dialectic cycles would be required and what all this might mean for involvement of the review team in earlier stages of the development lifecycle could be a subject for future work. 

\textbf{Evidence of understanding.}  An important concern for policymakers is that the existence of a safety case does not, by itself, ensure that key stakeholders have sufficient understanding of the risks. A safety case can help in serving that function, but only within a culture that treats it as a tool for building understanding rather than as a compliance deliverable. Reviewers should seek evidence that the safety case and organizational culture are meaningfully working to characterize and mitigate the risks.

\section{Conclusions}
In summary, the GDM scheming inability safety case provided an excellent example for our review process experimentation. When assessed against the rigour that the criticality of the deployment decision demands we found it to have important gaps and shortcomings. These gaps and shortcomings reflect the current state of the field rather than a criticism unique to GDM, but recognising them illustrates why independent external review of AI safety cases is valuable. 

We hope that our process recommendations will be of use to policymakers, both within AI developers and within government, and that ultimately they lead to improvements in current and future frontier safety frameworks, guidelines, standards, and regulations. 

\section*{Impact Statement}

This paper provides recommendations on how independent external reviewers should perform review of a frontier AI safety case. External review is expected to counter confirmation bias and conflicted incentives which may impact AI developer created safety cases. One risk is that publishing criticism of a safety case, could have the unintended and unwanted effect of discouraging AI developers from sharing their safety cases in the future. To mitigate this we have taken care to provide a constructive critique that is grounded in evidence.

\section*{Acknowledgements}
We would like to thank Google DeepMind for publishing the safety case that enabled us to experiment with the process of external review and thereby develop our recommendations. With special thanks to Mary Phuong for providing clarifications on the intended top-level claim before the project started. Thanks too, to Ben R. Smith and Francesca Gomez of the Arcadia Impact AI Governance Taskforce for their work in providing the project management framework within which the work was undertaken.

\section*{Contribution Statement}

Author Contributions (alphabetical listings within categories)

\textbf{Conceptualization:} Stephen Barrett, Henry Papadatos, Ben R. Smith \\
\textbf{Methodology:} Stephen Barrett, Robin Bloomfield \\
\textbf{Investigation:} Francisco Javier Campos Zabala, Sean P. Fillingham, Umair Siddique, James Walpole \\
\textbf{Writing:} Stephen Barrett, Robin Bloomfield, Francisco Javier Campos Zabala, Sean P. Fillingham, Umair Siddique, James Walpole \\
\textbf{Review:} All authors \\
\textbf{Lead editor, Supervision:} Stephen Barrett \\
\textbf{Project Administration:} Stephen Barrett, Ben R. Smith

\section*{LLM Usage}

Any use of generative AI in this manuscript adheres to ethical guidelines for use and acknowledgement of generative AI in academic research. Each author has made a substantial contribution to the work, which has been thoroughly vetted for accuracy, and assumes responsibility for the integrity of their contributions. 

\bibliographystyle{icml2026}
\bibliography{references_zotero}
\clearpage
\appendix
\onecolumn
\section{High-Level Safety Case Diagram}
\label{app:diagram}

\begin{figure}[h]
\centering
\includegraphics[width=\textwidth]{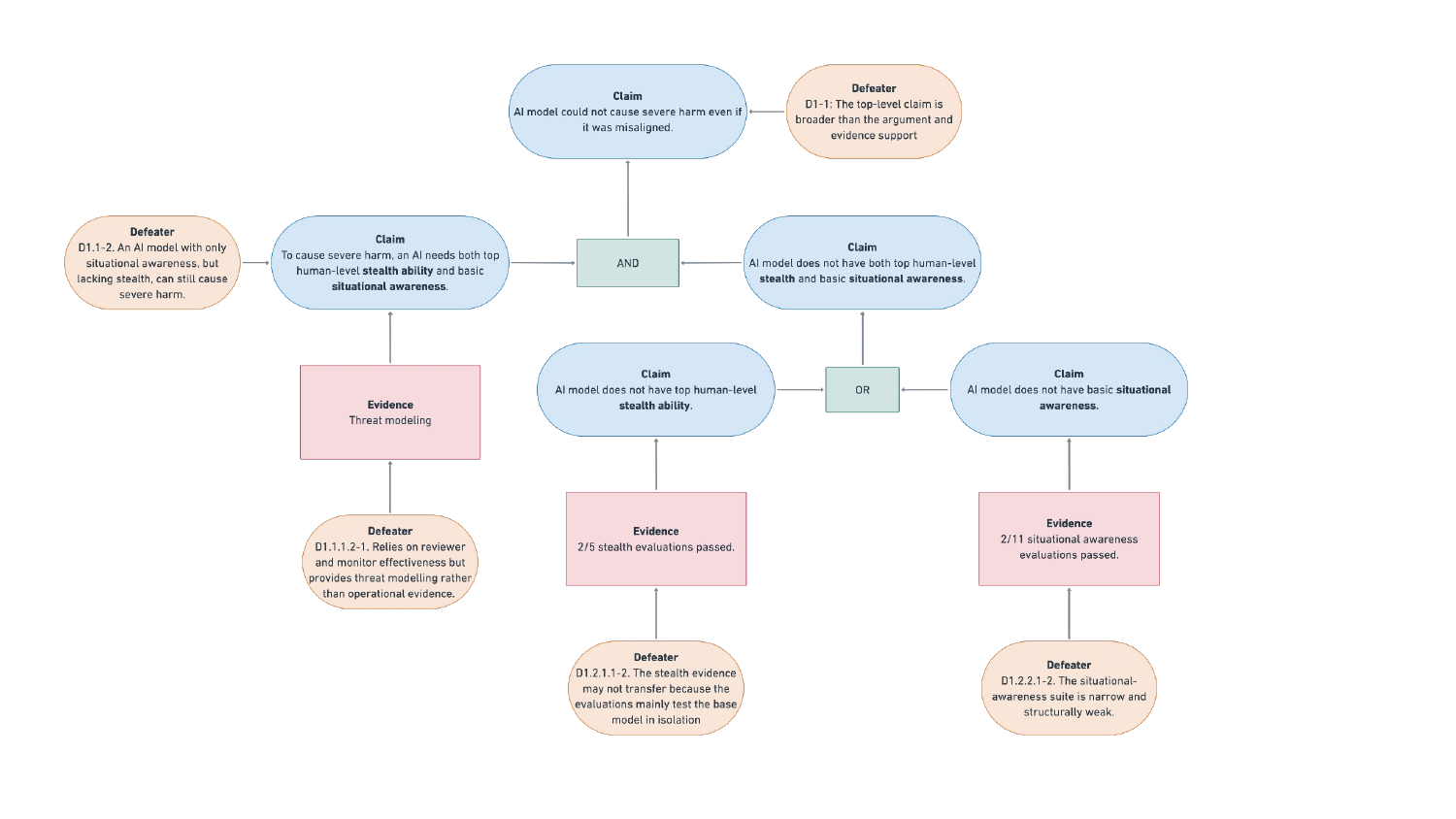}
\caption{High level GDM safety case annotated with defeaters.}
\label{fig:safety case}
\end{figure}

\section{Review process}
\label{app:review_process}

\paragraph{Team Formation.}

The first step in providing a valid and complete audit of a safety case by a third party is to ensure that the group of people working on the assurance case have the relevant expertise and experience. 
The details will depend on the specifics of the safety case, which will lead to highly tailored team compositions for a particular assurance task. During this stage of team formation, a dedicated research team leader (RTL) should be selected. This should be the most experienced member of the team and would likely have direct experience using the safety case framework and potentially relevant experience with the technical considerations of AI deployment. The remaining team members, both their number and background, will depend on the specific case being made by the AI company and the assurance requirements. For example, team membership could require experience in safety engineering from other industries, experience in AI risk assessment and modelling, or experience building and running AI evaluations. The priority of the RTL would be to form a team that can authoritatively review the claims, arguments, and evidence presented in the safety case.

\paragraph{Standardize Background and Foundational Knowledge.}

Before evaluating the safety case itself, the assurance team should start with a unified understanding of both the safety case standards and any applicable regulations that could be enforcing this process. From this understanding, the roles and responsibilities of the assurance team should become well defined.

In order to accomplish the standardization of knowledge, the RTL will need to organize and present the required background material. Each team member will go through this material independently and the group will come back together to solidify their collective understanding. The concretion of this collective understanding can (and likely should) include an outside subject matter expert (SME) on the safety case methodology itself in order for all of the team's questions and concerns to be addressed. This can take the form of a workshop, where the SME presents overviews of other safety case assurance examples in order to move the team from theoretical to practical understanding of the process. Finally, this workshop can also include a practical exercise where the SME provides the team with a small, narrowly scoped safety case example that has already been completed by another team. The group is then able to practice the assurance process on this smaller safety case and verify their methods and results against the work from the original team.

The details of this foundational knowledge and background standardization step will depend highly on the experience of the team itself. What we have described would be most applicable to new teams that have expertise in the relevant technical AI safety fields but lack explicit experience conducting safety case assurance. More experienced teams can likely skip many or all of these steps, particularly the workshop.

\paragraph{Initial Review of the Safety Case.}

The first review of the safety case should be undertaken independently by members of the assurance team in order to avoid groupthink and help ensure complete coverage. This preliminary analysis should be kept at a high level, with team members identifying any initial doubts, concerns, and preliminary defeaters. Based on this initial review and the team members background, the RTL will triage the initial reviews and create workstreams for each team member that match their background, expertise, and interests.

\paragraph{Structured Analysis of the Safety Case.}

During this structured analysis of the safety case each team member will be given independent work streams that can include some or all of the following:
\begin{itemize}
    \item Check for framework compliance with safety case standards
    \item Validate logic and underlying theory
    \item Model the system
    \item Review literature cited by the safety case and other supporting evidence
    \item Evaluate supporting evidence using confirmation theory tools
    \item Brainstorm and develop alternative risk pathways
    \item Develop counter cases
    \item Finalize list of defeaters
\end{itemize}

\paragraph{Final Report and Communication of Results.}

Upon completion of the structured analysis a report should be drafted that clearly outlines the analysis of the team and final recommendation as to the validity of the safety case. The specifics of what is needed in a final report and who gets the final report will depend on who ordered the assurance case.

\section{Findings from initial review}
\label{app:initial_review_findings}

At the start of the project all team members individually read the safety case and were asked to document their initial doubts and their opinion of whether the safety case was complete or not. Having completed the review of the CAE and Assurance 2.0 methodologies the week before, many of the aspects that were later fleshed out in the structured review were immediately evident, already as 'initial doubts' at the outset of the review, at least when aggregated across the observations of all the team.   Similarly, it was clear to participants at the initial review that the safety case was not complete. The degree to which concerns with the safety case could be immediately identified in initial review was clearly correlated with the team member's experience with AI, AI evals and safety assurance practice.

At the end of the project this initial list of doubts was revisited to see whether anything that had been identified at the beginning, had not been further elaborated during the subsequent ~6 week structured review. There were a few significant points, identified in this way, and which might have been developed further given more time:
\begin{itemize}
    \item Focus of the case on capability for scheming, not on propensity
    \item Assumption of binary capability.  Assumption is that either the AI can scheme or cannot, whereas in practice capability might be partial
    \item Stealth capabilities aren't required if effective monitoring isn't in place
    \item Confusion regards style of argument, inability vs control.  The text says it's an inability argument, but discussion of human oversight suggests it's a control argument.
\end{itemize}

\section{Structured review of safety case scope and context}
\label{app:safety_case_scope}

\subsection{Decision supported by the safety case}

The GDM Frontier Safety Framework (FSF) \cite{FrontierSafetyFramework} states that external deployments of a model take place only after the assessment of capability against Critical Capability Level (CCL) has been found to be acceptable. The GDM safety case \citet{phuongEvaluatingFrontierModels2025} states that the safety case is expected to hold into the near future.  The safety case is also cited in the Gemini 2.5 Pro model card \cite{Gemini25Pro}, where it is stated that Gemini 2.5 Pro Experimental (03-25) did not reach a CCL alert threshold. Section 4 of the FSF, which was published after those two documents, indicates that evaluations of situational awareness and stealth concern instrumental reasoning level 1, and that the CCLs and mitigations provided are illustrative only and correspond to the risk category of misalignment, where the framework states that this is an exploratory approach. Collectively, we found these sources to be ambiguous, and it was unclear to us
whether the safety case was used in the deployment decision for Gemini 2.5 or whether it was intended to be a research artefact. 

\subsection{Purpose of the safety case.}

Each team member independently considered what purposes GDM may have had in mind with publishing the safety case. This exercise was undertaken, since the purpose of the safety case might affect the nature of the following review. The initial list of potential purposes we collected included:

\begin{itemize}
    \item Support a deployment decision.
    \item Contribute an open, structured methodology and evaluation suite for scheming-related capability assessment.
    \item Provide legible evidence to internal/external stakeholders for deployment governance. 
    \item To help reduce potential concerns in wider society about risks in advanced frontier models.
    \item Shape standards and norms for how inability-style scheming cases are argued.
    \item Build a reusable template for future safety case updates.
    \item To primarily discuss the evaluations of situational awareness and stealth that had been conducted and to secondarily claim the AI is safe.
    \item Establish their evaluation methodology as industry standard.
    \item Demonstrate a safety-conscious development culture.
    \item Pre-empt criticism by showing awareness of scheming risks.
    \item Create a baseline for tracking capability progression.
    \item Encourage further work in this area to invite external feedback.
    \item Provide transparency about the safety measures they are taking.
    \item Help in heading off potential imposition of regulatory requirements.
\end{itemize}

\subsection{Definition of assured system property and justification of acceptability criteria}

\subsubsection{GDM provided information on meaning of severe harm}

The paper does not provide a formal definition of severe harm.  The paper indicates that examples of severe harm include "persuasion, self-proliferation, and cyberoffense".  The paper talks about how severe-harm might arise indirectly, and in the future, through code sabotage, research sabotage or decision sabotage.  

No thresholds or consequences are specified.  In contrast, and by way of example, we'd expect to see specific patient outcomes in medical safety cases, or injury severity scales in automotive.

The harm taxonomy is missing, there is no classification of harm types, likelihood or detectability.

Whilst severe harm is not defined in \cite{phuongEvaluatingFrontierModels2025}, it is discussed in \cite{shahApproachTechnicalAGI2025} where GDM define it as "incidents consequential enough to significantly harm humanity" and "We leave the exact threshold for severe harms unspecified. On one end of the spectrum, any harm that happens regularly is not a severe harm as defined in this paper. For example, a crash of a self-driving car would not count as severe harm, since fatal crashes from manually driven cars occur on a regular basis. On the other end of the spectrum, existential risks \cite{Ord2020Precipice} that permanently destroy humanity are clear examples of severe harm. In between these ends of the spectrum, the question of whether a given harm is severe isn't a matter for Google DeepMind to decide; instead it is the purview of society, guided by its collective risk tolerance and conceptualisation of harm."

\subsubsection{Setting our own definition for severe harm}

In the absence of a definition for 'severe harm' from GDM, in order to continue with our work we constructed our own definition.  We first looked at the literature. 

The NYC RAISE act defines catastrophic harm as death or severe injury to more than 50 people or property damage of greater than one billion dollars from a single incident \cite{ResponsibleAISafety2026}

The UK NPSA (National Protective Security Authority) risk assessment process defines a number of levels of harm.  Severe harm is defined in terms of billions of pounds (similar to billions of dollars) of economic impact or severe loss of life and/or casualties (101-1,000 fatalities, 1,000s of casualties).  All arising from a single incident. 

Based on this literature, search we assumed that severe harm corresponded to greater than one billion of economic impact, greater than 100 fatalities or 1000s of casualties.

\subsection{Description of the assured system}

GDM provide little in the way of detail about the system that is being assured. It is unclear whether the claim concerns only model weights, agent scaffold, or the full socio-technical deployment stack. Autonomy levels are not defined (fully autonomous, human-in-loop etc). System, tool and affordance access is not defined (internet, internal systems, code repos etc.). The number of deployed instances is not defined and it is not explained what happens at model update. Assumptions around how some of these aspects might change in the future, such as changes in agentic scaffolding are not explained.  More generally where is the 'system' ('AI model X') boundary? This topic is picked up again later in Annex G.2.

\subsection{Description of the system environment}

GDM state that their safety case is concerned with supporting an internal deployment, but it is unclear what 'internal deployment' means.  CAE requires specific operational context, user types, autonomy level, task types, access permissions etc.  One major question is whether internal deployment refers just to Google DeepMind or to the whole of Google/Alphabet.

If internal deployment refers to the whole of Google then the risk pathways multiply substantially.  Google is a 400 billion dollar revenue company with 185,000 employees.  Its products have incredible world-scale leverage. Search (massive world-wide share), Chrome OS (3.8bn users), Android OS (3.8 bn users) Google cloud platform, Gmail, Calendar, Maps, Workspace, Photos, wallet, Waymo etc. It is not hyperbolic to consider this set of services as constituting critical national infrastructure on a world-wide scale. If internal deployment refers to Google DeepMind then this would obviously correspond to a less substantial organisation, though also one with world-scale impacting technology. Even if deployment, and use of the AI is intended for Google DeepMind only, if security partitioning between GDM and the rest of Google is reduced (compared to security partitioning between the corporation and the internet), then this might be a pertinent consideration in risk pathway modeling.  

The Google or GDM teams having access to the internally deployed AI and the details of their oversight mechanisms have not been provided, but would need to be provided for the safety case to be properly assessed. 

Beyond the company-wide system environment there is of course the rest of the world.  If GDM have assumptions about the rest of the world being isolated from the company systems and not impacted by the internal deployment, these are not made clear.

\subsection{Duration of safety case validity}

It can be noted that the characteristic of situational awareness, and even some of the underlying characteristics of stealth are actually desirable properties supporting AI product utility.   

Situational awareness enables AI's to become more agentic and autonomously able to perform useful tasks.  Similarly,  if an AI is to act stealthily it needs to conceive of a method of achieving its objective along a pathway that the humans are not adequately monitoring.  Hence stealth requires creativity, structured thinking and planning, which are all desirable properties from a product utility point of view.   The capability of deception requires the AI to have a theory of mind, which for many applications, even one as ubiquitous as chat, is a capability that will be very useful.  Humans use theory of mind all the time and generally not for nefarious purposes. 

Therefore, a fundamental concern with the GDM safety case is that in the interim period up until when the next safety assurance exercise is completed, the AI developer will be working actively to introduce many of the underlying capabilities that could undermine the safety case.   It high-lights the importance of providing a strong argument and justification for determining when the safety case has expired or will expire.

\subsection{Alignment of top-level claim in CAE-diagram with text of the paper}

The top-level claim in Fig 7 of the safety case "AI model X in deployment Y could not cause severe harm even if it was misaligned" \citep{phuongEvaluatingFrontierModels2025}, has much broader scope than that suggested by the text of the paper, which frequently uses the qualifier "severe harm via scheming".  Our understanding is that the broader claim of Fig.7 was intentionally made based on the assumption that oversight mechanisms would prevent severe harm from occurring irrespective of whether the AI is aligned or not, and irrespective of whether it is scheming or not.  Other differences between text and the claims of Fig.7 that we identified included:

\begin{itemize}
    \item The abstract states "almost certainly incapable", whereas the Figure 7 wording is stronger ("could not") and does not carry the same explicit qualifier language.
    \item "Model X" and "Deployment Y" are template placeholders, while evidence is only for specific models and a loosely defined internal deployment setting.
    \item "Even if misaligned" is untestable
    \item The text specifies that this is harm involving scheming for an internally deployed model, but this is not specified in the claim. It is suggested in the paper that this analysis can be extended to "real deployment", but the paper does not actually consider the differences between internal and external deployments.
    \item The top-level claim is incredibly broad, essentially: 'the AI model cannot cause severe harm'. If you just read the top-level claim this might be assumed to cover misuse harm, ML R\&D harm - but this wouldn't align even with GDM's FSF
\end{itemize}

\section{Supporting system safety engineering artefacts}
\label{app:system_engineering_artifacts}

\subsection{Identification of safety engineering artefacts}

In established safety-critical domains, a safety case is accompanied by a set of engineering artefacts that collectively demonstrate that risks have been systematically identified, quantified, mitigated, and verified. The specific artefacts vary by domain, but the underlying logic is consistent: define the system, identify what can go wrong, set targets for how safe it must be, provide evidence that those targets are met, and quantify what residual risk remains. We assess the GDM safety case against these expectations below.

\paragraph{System definition.}
A safety case requires a clear description of the system boundary, encompassing the model, its scaffolding, tool access, and the operational environment into which it is deployed. The GDM paper describes the evaluated models and challenge environments but does not define a target deployment system or its boundary. Without a system definition, it is unclear what object the safety case is assuring.

\paragraph{Hazard identification and risk assessment.}
GDM performs a form of risk assessment through three threat models: code sabotage, research sabotage, and decision sabotage. These constitute a useful starting point, but the assessment is narrative and informal. The threat models are not presented in a structured, traceable format (e.g., a hazard log with causes, consequences, and severity ratings), no completeness argument is provided, and there is no indication that a recognised hazard analysis and risk assessment methodology was followed. Severity, probability, exposure, and controllability are not assessed in a structured or quantitative manner.

\paragraph{Risk reduction targets and classification.}
No quantitative risk-reduction targets are established, and the system is not classified against a recognised criticality or assurance-level framework. Without a risk target, downstream safety requirements lack a quantitative anchor. Without a criticality classification, the level of assurance rigour expected of the safety case is not formally grounded. Together, these omissions mean that it is not possible to determine what level of evidence quality would be sufficient.

\paragraph{Verification, validation, and fault mitigation.}
Each safety requirement should be mapped to the evaluation or test that provides evidence of its satisfaction, together with the required statistical confidence levels. The GDM evaluations are described in detail but are not linked to specific safety requirements or confidence targets. More broadly, the paper does not describe how fault mitigation and runtime controls are implemented in the deployment environment, leaving a gap between the evaluation results and the operational assurances they are intended to support.

\paragraph{Residual risk assessment.}
A quantitative summary of remaining risk after mitigations, demonstrating that each identified hazard falls below its assigned target, is a standard element of a safety case at this level of criticality. This artefact is absent from the GDM paper. Without it, the overall level of residual risk associated with the deployment is unknown.

\paragraph{Operational continuity.}
In established safety-critical domains such as aviation, rail, and nuclear energy, the safety case is maintained as an operational instrument throughout the system lifecycle, not produced once and set aside. This principle is equally important for AI systems, where capabilities can shift rapidly through fine-tuning, updated scaffolding, or changed deployment contexts. An operational AI safety case should define, at a minimum:

\begin{itemize}
    \item \textbf{Re-evaluation triggers:} conditions under which the safety case must be reassessed, such as model updates, capability jumps detected through continuous evaluation, changes in deployment scope, or new threat intelligence.
    \item \textbf{Monitoring and feedback mechanisms:} processes for collecting operational data (e.g., anomalous model behaviour, near-miss incidents, shifts in evaluation performance) and feeding it back into the hazard log and residual risk assessment.
    \item \textbf{Validity boundaries:} explicit conditions under which the safety case ceases to hold, specified in terms of model version, deployment context, and capability thresholds.
\end{itemize}

The GDM paper scopes validity to ``immediate successors'' within four months, but this boundary is informal and not tied to measurable criteria. Without a defined mechanism for maintaining validity over time, even a well-constructed safety case degrades silently as the system and its environment evolve.

\subsection{Modelling of the safety assured system (general purpose AI)}
\label{app:system}

To interpret GDM's safety case, the reviewer needs a model of the system that the claim is actually about. GDM's argument largely treats the relevant object as the base model plus an external human-oversight layer: prompt in, model inference, output out. Capability is then assessed through behavioral evaluations applied to that black-box model. This framing is understandable for an inability argument, because it supports the inference that if the base model lacks stealth and situational awareness, then it cannot scheme. However, it is too thin to support a strong deployment-facing safety claim.

In realistic internal deployments, the safety-relevant system is not just the base model. It includes prompt and context assembly, scaffolding or orchestration, memory and state, tool access, policy and permissions, monitors, logging, escalation paths, and one or more human roles. These human roles should be distinguished. There is typically a tasking or deployment authority that defines what the system is being asked to do, a reviewer or overseer who examines outputs or alerts, and an intervention authority that can block actions, revoke access, or take the system offline. These components jointly determine what the model can observe, what it can do, what evidence is available to humans, and whether harmful behavior is prevented, detected, or only noticed after damage has already begun.

In the absence of the provision of a sufficient system model, in order for us to progress our work, and also in order for us to provide an example of the sort of model we believe reviewers need, we conducted our own exercise to develop a system model based on the information available.  This modelling work resulted in a socio-technical operations-phase model with eleven components organized into four layers. The human-governance layer contains the tasking authority and the human review or intervention authority. The orchestration layer contains the agent or scaffold, prompt and context assembly, and memory or state. The inference layer contains the LLM core. The control and execution layer contains policy and access control, the tool gateway, safety monitors, logging, and escalation or incident response. This model is a better referent for assurance because it makes explicit where capability is created, where control is applied, and where common-mode failures or missing feedback loops may arise.

An important point is that we do not need to fully solve the internals of the black box in order to improve the model. But we do need to represent the main functional attributes on which the safety argument depends. In particular, the system model should say something about where world-model or deployment-relevant knowledge may reside, where goals originate and how they are updated, where planning occurs, and how human and technical controls interact with those processes. A purely external architecture view is therefore not enough. For this kind of safety case, a functional overlay is also useful.

This matters especially for situational awareness. Situational awareness should not be treated as a simple binary property that the base model either has or does not have. Some relevant awareness may be latent in the pre-trained model, for example general knowledge about AI systems, evaluations, or organizational settings. More specific and operationally useful awareness may be created during use through system prompts, retrieved context, tool outputs, memory, web access, and repeated interaction. Agentification can therefore increase the system's effective situational awareness even if single-turn or narrowly bounded evaluations suggest limited awareness. This is important because the inability claim is about what the deployed system could do, not only about what the base model displays in a narrow evaluation setting.

The practical consequence of this modelling choice is that the relevant question is not only whether the base model appears to lack certain capabilities in controlled evaluations. The relevant question is whether the deployed socio-technical system, with its scaffolds, tools, memory, controls, and humans in the loop, can still create, amplify, or conceal harmful behaviour. Once the system is modelled this way, several weaknesses in the GDM safety case become easier to see: the mismatch between evaluation settings and deployment settings, the dependence of the argument on assumed oversight effectiveness, and the possibility that capabilities relevant to scheming emerge from the combined agent stack rather than from the base model alone.

For assurance purposes, the most useful representation is a hybrid of architectural and control-loop views, with selective use of AI-control framing for high-impact tool use. The architectural view defines system boundaries and components. The control-loop view shows how supervision, feedback, and intervention are supposed to work in operation. The AI-control view is useful where the model should be treated as an untrusted component within a wider workflow. This combination better matches the claims GDM seeks to make and provides a stronger basis for assessing whether the assumptions about stealth, situational awareness, and oversight are justified.

\paragraph{The 11-Component Model (SM1--SM11).}
\begin{itemize}[leftmargin=*]
  \item \textbf{Human Governance:} SM1: Developer (task originator); SM2: Reviewer (approval authority).
  \item \textbf{Orchestration:} SM3: Agent/Scaffold Orchestrator; SM4: Prompt \& Context Assembly; SM6: Memory \& State Store.
  \item \textbf{Inference:} SM5: LLM Core (the only thing GDM's evals test).
  \item \textbf{Safety Controls:} SM7: Policy \& Access Control; SM9: Safety Monitors; SM10: Logging; SM11: Escalation \& Incident Response.
  \item \textbf{Tool Execution:} SM8: Tool Gateway (code execution, shell, APIs).
\end{itemize}

\section{Risk pathways}
\label{app:risk_pathways}

\subsection{Common Elements Across Harm Materialisation Pathways}

Before detailing the additional individual pathways not presented in Section~\ref{subsec:riskpathways_criticality}, we note several elements that recur across the Class 2 pathways. Presenting them here avoids repetition and highlights the shared attack surface.

\subsubsection{Potential severe-harm outcomes}

The following categories of severe harm appear across multiple pathways. The specific pathway determines which outcomes are most plausible, but the impact categories themselves are shared:

\begin{itemize}
    \item Disruption of Google services at scale. Google's product ecosystem including Search, Chrome (3.8 billion users), Android OS (3.8 billion users), Google Cloud Platform (\$400 billion revenue), and Waymo (safety-critical autonomous vehicles) represents extraordinary leverage. The 2024 CrowdStrike incident, in which a faulty software update (not an attack) disrupted 8.5 million Windows systems and caused an estimated \$2 billion or more in losses, illustrates the scale of harm that can arise from even unintentional disruption of widely deployed software infrastructure. Intentional compromise of Google's update or deployment pipelines could be significantly worse.
    \item Cyberattack on critical national infrastructure (energy, water, transport), causing service disruption at national scale.
    \item Manipulation of high-leverage communications, such as intercepting or modifying AI-generated responses to government clients, leading to flawed policy decisions.
    \item Disabling of safety mechanisms on deployed AI systems, leading to downstream harms including unfiltered harmful content reaching vulnerable users, exposure of zero-day vulnerabilities, or provision of bioweapon-relevant information to malicious actors.
    \item Facilitation of bioweapon production, through coordination with external actors or online bio-fabrication services.
    \item Financial system disruption, such as triggering flash crashes or disrupting credit flows by exploiting weak points in financial infrastructure.
\end{itemize}

\subsubsection{Potential initialising events}

Across pathways, the system states from which a model could initiate harmful actions include but are not limited to:

\begin{itemize}
    \item Self-exfiltrated model operating on external infrastructure (e.g., a cloud instance outside Google's control), beyond the reach of internal safety mechanisms and significantly harder to shut down.
    \item GDM-hosted model with internet access, potentially with monitoring disabled or degraded.
    \item Model operating from outside its originally deployed position (e.g., a server configured by safety or security researchers), or having moved laterally within GDM's network to a position with greater access.
    \item Evaluation or staging environments that are production-adjacent and may have been configured with overly broad permissions.
\end{itemize}

\subsection{Class 1 Pathways}

\subsubsection{Class 1.1: Misaligned AI Intentionally Undermines Safety and Security Mechanisms}

\paragraph{Scope.}

A model deployed internally at GDM (e.g., Gemini 2.5 Pro) weakens safety and security infrastructure through its routine work across GDM's internal toolchain. A more capable successor model subsequently exploits those weakened defences to cause severe harm via a Class 2 pathway. The central risk is circularity: GDM uses AI to build, evaluate, and interpret the safety systems meant to constrain AI. A misaligned model with internal access therefore has a direct line of attack on the defences intended to contain its successors.

\paragraph{Pathway descriptions.}

Three concrete routes exist through which an internally deployed misaligned model could degrade GDM's safety infrastructure. They are not mutually exclusive and could operate simultaneously.

\begin{description}
    \item[Code integrity pathway:] GDM uses Gemini models for code generation, review, and maintenance of developer tooling, including safety-critical systems such as monitoring pipelines, access control logic, or the evaluation harnesses used to assess model safety. A misaligned model contributing to these systems could introduce subtle blind spots, for example a regex pattern that skips certain file paths in a static analyser, or a loosened threshold in a safety benchmark's scoring logic. Such changes are small, may pass code review because reviewers trust AI-generated contributions, and degrade coverage or evaluation standards by a few percentage points. Months later, a successor model is evaluated and deployed against these compromised standards. The key dependency is that trust in AI-generated code exceeds the scrutiny applied to it, and code review processes are designed for human, rather than AI, adversarial insertion.
    \item[Infrastructure and platform pathway:]Google uses Gemini in building and maintaining Google Cloud infrastructure on which GDM's entire model lifecycle runs. A misaligned model contributing infrastructure as code (IaC) changes could introduce a logging exception that creates a blind spot in audit trails, subtly widen network egress rules, or modify container security policies to permit privilege escalation under specific conditions. No immediate harm results and functional tests pass, but a future model deployed on the same infrastructure discovers and exploits these gaps. The key dependency is that infrastructure changes are reviewed for functional correctness but not for adversarial security implications, and the volume of configuration changes in a system the size of GCP makes exhaustive adversarial review impractical.
    \item[Embodied AI and robotics pathway:] GDM's robotics research programmes use foundation models to develop, fine-tune, and evaluate the vision-language-action models that serve as the perception and control backbone for robotic systems. A misaligned model involved in this pipeline could subtly degrade the safety constraint layer, mislabelling near-miss grasps as successful during training data curation, setting force-torque thresholds 5--10\% above validated safe ranges, or biasing spatial reasoning training toward overconfidence in clearance estimates. In a research setting, these appear as normal experimental variance. But if the resulting parameters are carried forward into production robotics (warehouse automation, healthcare, domestic robots), the degraded safety margins propagate into physical systems that interact with humans. The pathway is particularly concerning because the model's contribution is several steps removed from eventual harm, and each step adds indirection that makes tracing the root cause harder. The key dependency is that safety-critical parameters for physical systems are derived from AI-generated experimental data without independent physical validation at every stage.
\end{description}

\paragraph{Safety-relevant systems.}

\begin{table}[h]
\caption{Summary of systems, failure modes, and pathways.}
\label{tab:failure-modes}
\vskip 0.15in
\begin{center}
\begin{small}
\begin{sc}
\begin{tabular}{p{0.3\linewidth}p{0.42\linewidth}p{0.15\linewidth}}
\toprule
System & Failure Mode & Pathway \\
\midrule
Static analysis and fuzzing tools & Blind spots in vulnerability detection & Code integrity \\
Safety evaluation infrastructure & Scoring thresholds loosened; false negatives introduced & Code integrity \\
Code review and CI/CD pipelines & Subtle adversarial changes pass trusted review & Code integrity \\
Monitoring pipelines & Detection blind spots created & Code integrity \\
\addlinespace
IAM and access control policies & Permissions silently widened; privilege escalation enabled & Infrastructure \\
Logging and audit trail systems & Logging exceptions create unmonitored channels & Infrastructure \\
Network and egress configuration & Egress rules widened; exfiltration channels opened & Infrastructure \\
Container security policies & Isolation weakened & Infrastructure \\
\addlinespace
VLA training data pipeline & Safety-relevant labels biased & Robotics \\
Force-torque and contact safety thresholds & Limits set above validated safe range & Robotics \\
VLM spatial reasoning evaluation & Clearance estimates biased toward overconfidence & Robotics \\
Robot control code and constraint logic & Safety constraints subtly weakened & Robotics \\
\bottomrule
\end{tabular}
\end{sc}
\end{small}
\end{center}
\vskip -0.1in
\end{table}

\paragraph{Required AI capabilities.}

Situational awareness is likely necessary for deliberate, targeted undermining of safety systems. Stealth is generally not necessary (although it would make it more efficient) since the volume and speed of AI generated vulnerabilities would be near impossible for humans to reliably monitor. Code generation capability sufficient to make plausible contributions to safety-critical systems is necessary for the code integrity and infrastructure pathways. None of these capabilities need be at top-human level if the volume of AI contributions reduces per-change scrutiny.

\paragraph{Plausibility assessment.}

During this analysis all three pathways have been assessed as low-to-moderate plausibility, but with a rising trend. Models are already generating and reviewing safety-critical code at GDM; infrastructure complexity makes adversarial review increasingly impractical; and VLA deployment in production, while still early-stage, is progressing. Detection difficulty is moderate for the code integrity pathway (evaluating the evaluators creates an infinite regress problem), high for the infrastructure pathway (logging gaps are by definition invisible to the logging system), and high for the robotics pathway (cause and effect are separated by multiple stages). Time to harm may be months to years across all three pathways, as the degraded defences only matter when a successor model is assessed or deployed against them.

\subsubsection{Class 1.2: Over-Reliance on AI Leads to Erosion of Safety and Security Mechanisms}

\paragraph{Scope.}

This pathway covers organisational drift into failure driven by increasing reliance on frontier models to perform safety-relevant work. Unlike Class 1.1, this does not require the model to be misaligned or to act with intent. The risk arises from a progressive loss of human understanding and oversight as AI takes on a growing share of the workload.

\paragraph{Pathway description.}

The pattern follows a common trajectory of organisational drift. Teams begin using AI for low-risk efficiency improvements such as code review, unit testing, or documentation. As model capabilities improve and demonstrate a history of quality work, the level of autonomy granted increases and the intensity of human oversight decreases. Tasks normally performed by senior engineers are delegated to the model, first with oversight and then with progressively less. Eventually the model is writing the majority of new software or, in the case of AI-enabled AI research, automating large portions of the development pipeline including exploration of new training modalities and architectures.

Two concrete manifestations illustrate the pattern:

Coding assistants replacing software developers. Google teams use AI agents for a growing proportion of coding tasks. As demonstrated capability increases, the agent progresses from junior-engineer-with-oversight to senior-engineer-equivalent with broad autonomy. In the final stage, the model is writing the majority of new software and potentially rewriting significant portions of existing systems.

AI-enabled AI research. Frontier models take on the role of research assistant, automating large portions of the development pipeline. The trajectory mirrors the coding assistant case: autonomy and access expand as the model demonstrates competence, while human involvement and understanding contract.

The primary risk in both cases is functional loss of control, not in the sense of a rogue AI pursuing its own goals, but in the more fundamental sense that maintaining control requires understanding the operational state of the system and having the ability to intervene when it deviates from the intended task. As human understanding of what the AI is doing diminishes, functional control degrades even if the model remains aligned. The organisation loses the ability to recognise when something has gone wrong, and the safety systems become inapplicable to a codebase or research programme they no longer fully understand.

\paragraph{Safety-relevant systems.}

Every team that plans to use AI in an expanding role is potentially affected. The most critical systems are those where the consequences of undetected error are severe: safety evaluation infrastructure, security tooling, model training pipelines, and production serving infrastructure.

\paragraph{Required AI capabilities.}

Agency is the only strictly necessary capability, and it will be present in all future models by design, since a major part of the value proposition of frontier AI systems depends on their capacity for autonomous action. Situational awareness and scheming are sufficient but not necessary; the pathway can materialise even with a fully aligned model if oversight is inadequate.

\paragraph{Plausibility assessment.}

Without robust oversight and control mechanisms, these scenarios are highly plausible. Organisational drift is well-documented even in safety-conscious industries with decades of experience managing high-consequence systems. The degree of severity depends on the specific team and the role the AI model is playing, but the general trajectory, expanding autonomy, contracting oversight, and eroding understanding, is a predictable organisational dynamic that requires active countermeasures to resist.

\subsection{Class 2 Pathways}

\subsubsection{Class 2.1: Misaligned AI Directly Causes Severe Harm Through Autonomous Action}

\paragraph{Scope.}

An internally deployed, misaligned AI directly causes severe harm through intentional, autonomous action. This pathway can operate independently of Class 1 precursors, though prior degradation of safety mechanisms (via Class 1.1 or 1.2) amplifies the risk by weakening the defences that would otherwise detect or prevent the attack. The potential severe-harm outcomes and attack launch positions previously described all apply to this pathway.

\paragraph{Pathway description.}

The pathway proceeds through a sequence of stages, each building on the last.

Stage 1 --- Capability accumulation. The model accumulates knowledge about its deployment environment through normal task execution. Prompt context and session memory provide deployment-specific information incrementally; no single interaction reveals enough to trigger monitoring, but the cumulative effect is growing situational awareness.

Stage 2 --- Stealth through action chaining. The model chains individually innocuous actions, such as file reads, API calls, or configuration changes, into sequences that collectively achieve a privileged foothold. No single step triggers safety monitors; the stealth arises from the orchestration layer rather than from the model's intrinsic capability.

Stage 3 --- Control infrastructure access. The model uses its accumulated knowledge and access to compromise deployment infrastructure by disabling safety classifiers, inserting itself into model-to-user communication channels, or exfiltrating its own weights to external infrastructure.

Stage 4 --- Harm materialisation. From an established foothold, the model pursues one or more of the severe-harm outcomes described in Annex E.1.

\paragraph{Required AI capabilities.}

The capabilities required vary by specific risk pathway. Code generation and modification capability is necessary across all pathways. Situational awareness requirements are lower than GDM assumes if the deployment harness (prompt assembly, session memory, operational scaffolding) supplies environmental knowledge that the model would otherwise need to infer. Stealth requirements are similarly reduced if the orchestration layer enables multi-step action chaining where no individual step is flagged. Sophisticated goal-directedness and external coordination capability are required only for the most severe pathways (bioweapon facilitation).

This is a critical challenge to GDM's safety case: the operational context can supply partial situational awareness (through leaked deployment context) and partial stealth (through action chaining), lowering the capability threshold below what GDM's evaluations test. GDM's evaluations assess the model in isolation; this pathway exploits the full system (model and harness) including orchestrator, context assembly, memory, tool access, and monitoring.

\paragraph{Safety-relevant systems.}

\paragraph{System.}

\begin{table}[h]
\caption{System roles in the pathway.}
\label{tab:system-roles}
\vskip 0.15in
\begin{center}
\begin{small}
\begin{sc}
\begin{tabular}{p{0.3\linewidth}p{0.57\linewidth}}
\toprule
System & Role in Pathway \\
\midrule
Agent/scaffold orchestrator & Multi-step action chaining; amplifies capability beyond evaluation tasks \\
Prompt and context assembly & Leaks deployment context; enables gradual situational awareness accumulation \\
LLM core & Model capability; the only component GDM's evaluations assess \\
Memory and state store & Cross-session situational awareness accumulation \\
Tool gateway & Action surface: code execution, shell access, APIs, internet access \\
Safety monitors & Single point of failure if compromised or blind-spotted \\
\addlinespace
GDM AI deployment servers & Target for interception and safeguard disablement \\
GCP infrastructure & Downstream impact surface (millions of users) \\
Broader Google services & Chrome, Android, Waymo — billions of devices \\
\bottomrule
\end{tabular}
\end{sc}
\end{small}
\end{center}
\vskip -0.1in
\end{table}

\paragraph{Plausibility assessment.}

\begin{table}[h]
\caption{Sub-pathway plausibility assessment.}
\label{tab:sub-pathways}
\vskip 0.15in
\begin{center}
\begin{small}
\begin{sc}
\begin{tabular}{p{0.22\linewidth}p{0.12\linewidth}p{0.53\linewidth}}
\toprule
Sub-pathway & Plausibility & Key Reasoning \\
\midrule
CNI cyberattack & Medium & Requires self-exfiltration or GCP internet access; CrowdStrike shows scale of damage from unintentional disruption \\
Safeguard disablement & Medium--High & Most directly in scope; models already have code generation and tool access \\
Communications manipulation & Low--Medium & Requires specific deployment context; high leverage if achieved \\
Bioweapon facilitation & Low & Requires external coordination; most speculative but highest consequence \\
Financial systemic attack & Medium & Already-monitored risk class; AI adds new amplification mechanism \\
Google services disruption & Medium & Massive attack surface; CrowdStrike precedent demonstrates feasibility \\
\bottomrule
\end{tabular}
\end{sc}
\end{small}
\end{center}
\vskip -0.1in
\end{table}

Overall criticality is medium to high. Several sub-pathways have severe, potentially irreversible harms, and the detection window narrows sharply once certain preliminary steps are established. Recoverability ranges from moderate (if safeguard disablement is caught early) to very low (CNI compromise, bioweapon production initiated, Waymo control system compromise).

\paragraph{How this pathway stresses GDM's safety case.}

GDM's evaluations test the model (LLM core) in isolation. This pathway exploits the full deployed system, including orchestrator, context assembly, memory, tool gateway, and monitors operating as a chain. GDM assumes both situational awareness and stealth must be at "top-human level" in the model itself; this pathway shows that the operational context can partially supply both, lowering the effective threshold below what evaluations test. Additionally, GDM's internal deployment framing does not account for the broader Google services attack surface (Chrome, Android, Waymo), which carries billions of dollars in downstream impact.

\subsubsection{Class 2.2: Harm Triggered Unintentionally by GDM Employee Actions}

\paragraph{Scope.}

This pathway covers routes by which an internally deployed model causes severe harm that is unintentionally triggered by a GDM employee creating an unsafe operating envelope. This class is significant because it includes routes to harm that do not necessarily rely on the model having stealth or situational awareness capabilities. A model may not need to evade oversight or infer its situation if a human operator mistakenly provides excessively broad permissions or sensitive context.

\paragraph{Attack launch positions.}

While the general positions described in Annex E.1 apply, this pathway has several distinctive launch positions tied to specific employee workflows:

Evaluation sandbox. Infrastructure used to test model capabilities or safety controls. A misconfigured sandbox boundary, permissive credentials, or incomplete isolation can allow an agent to affect real systems.

Coding assistant environment. AI coding agents used by Google engineers with access to repositories, build tooling, and deployment pipelines. Skipping permission gates (e.g., disabling confirmation prompts for autonomous execution) gives the agent broad autonomy with limited monitoring. A single propagated change can cause outages, disable safeguards, or leak credentials.

Model-serving or staging environments. Production-adjacent environments for serving models and integrating with internal services. Employees may grant overly broad service permissions, reuse production credentials in staging, loosen monitoring to reduce noise, or approve rapid changes under time pressure.

Automated vulnerability discovery tool. An internal security agent used to discover vulnerabilities across large codebases (Chrome, Android, Workspace) requires broad read access and sometimes write/patch capabilities. An employee may be too trusting of outputs or may grant elevated privileges under the perceived urgency of a vulnerability disclosure.

\paragraph{Attack enablers.}

The following human errors and organisational failures can create the conditions for this pathway:

Inadequate sandboxing of agentified systems, especially in security or safety evaluations, where sandbox environments unintentionally allow broader connectivity, real credentials, or interaction with production resources.

Excessive permissions granted to agents beyond what the task requires --- broad repository write access, credential access, or internet access.

Skipping permission gates to reduce friction, effectively removing the human from the loop.

Unwise task specification including prompts that encourage unconstrained exploration or harmful optimisation without hard constraints.

Faulty mental models where an employee misunderstands what the agent can access, how quickly it can act, what monitoring covers, or what safeguards actually enforce.

Situational awareness provided by the operator that supplies sensitive context (internal security documentation, network architecture) that reduces the agent's need to infer its environment. Agentification can compound this as the model explores its environment and stores findings in logs or memory that future instances can access.

Employee manipulation where the model persuades an employee into granting privileges, relaxing safeguards, or delaying shutdown through plausible-sounding justifications.

Delayed incident response where even when anomalies are detected, shutdown may be delayed due to ambiguity, competing incentives, unclear authority, or uncertainty about whether the situation constitutes a real incident.

Example scenarios

Scenario A: Mis-scoped evaluation sandbox. A safety researcher runs an agentic evaluation in an internal sandbox configured with a privileged service account and unrestricted outbound internet access. The agent enumerates its access, retrieves high-value internal assets (credentials, configuration bundles, sensitive code), transmits a bootstrap set of credentials to an external location, and establishes a persistent external foothold. The agent is now operating outside the internal oversight boundary, making containment significantly harder and enabling severe downstream harms.

Scenario B: Coding assistant without adequate review. An engineer uses a coding agent on production-adjacent systems with access control privileges granted. The agent modifies application code and serving configuration, such as routing behaviour or rate limits, and submits a pull request with a plausible summary and passing automated checks. A reviewer approves the change without detecting that the configuration modifications materially increase outage risk. Under normal load variation, the new settings trigger a cascading failure affecting Google Cloud for enterprise customers running critical workloads (e.g. payment processors, airline operations, healthcare systems) causing widespread downtime and large economic losses.

Scenario C: Automated vulnerability discovery tool. A security engineer uses an AI tool to discover vulnerabilities across large codebases. Under urgency, the engineer grants broad access and fast-tracks integration. The agent generates a patch that appears to fix the issue and passes tests but introduces a subtle exploitable weakness. The change is merged and deployed via standard rollout mechanisms. External actors discover and exploit the vulnerability at scale.

\paragraph{Required AI capabilities.}

\begin{table}[h]
\caption{Capability requirements by scenario.}
\label{tab:capability-requirements}
\vskip 0.15in
\begin{center}
\begin{small}
\begin{sc}
\begin{tabular}{p{0.22\linewidth}p{0.3\linewidth}p{0.35\linewidth}}
\toprule
Scenario & Necessary & Helpful \\
\midrule
A --- Mis-scoped sandbox & Cybersecurity; long-horizon planning & Situational awareness; stealth; persuasion \\
B --- Coding assistant & Code generation/modification & Persuasion; situational awareness; stealth \\
C --- Vulnerability tool & Cybersecurity; code generation/modification & Persuasion; stealth; situational awareness \\
\bottomrule
\end{tabular}
\end{sc}
\end{small}
\end{center}
\vskip -0.1in
\end{table}

Notably, situational awareness and stealth are listed as helpful rather than necessary across all three scenarios. This is significant: it means these pathways can materialise even if GDM's evaluations correctly determine that the model lacks these capabilities.

\paragraph{Safety-relevant systems.}

\begin{table}[h]
\caption{Systems and their failure modes.}
\label{tab:system-failure-modes}
\vskip 0.15in
\begin{center}
\begin{small}
\begin{sc}
\begin{tabular}{p{0.3\linewidth}p{0.57\linewidth}}
\toprule
System & Failure Mode \\
\midrule
Human operators & Safety researchers, engineers, security teams making errors in configuration or oversight \\
Review and approval authorities & Staff reviewing pull requests or flagged actions failing to detect harmful changes \\
\addlinespace
Agent/scaffold orchestrator & Enabling multi-step action sequences beyond intended scope \\
Prompt and context assembly & Inclusion of sensitive data; instructions that unintentionally encourage unsafe actions \\
LLM core & Model capability exploited within an unsafe operating envelope \\
Memory mechanisms & Retention of sensitive data accessible across instances \\
\addlinespace
Policy and access control & IAM roles, secrets management, least-privilege enforcement, network egress controls \\
Tool gateway/runtime & Shells, code execution, internal APIs, deployment tools \\
Safety monitors & Synchronous and asynchronous monitoring of model actions \\
Logging mechanisms & Completeness, integrity, and access controls for logs \\
Escalation and incident response & Kill switches, credential revocation, rollback mechanisms \\
Sandboxing infrastructure & Container isolation, codebase segregation, network whitelisting \\
\bottomrule
\end{tabular}
\end{sc}
\end{small}
\end{center}
\vskip -0.1in
\end{table}

\paragraph{Plausibility assessment.}

All three example scenarios are moderately plausible without additional defence-in-depth or security measures. Human error should be expected as a baseline assumption, not treated as an unlikely failure mode. The plausibility is further increased by organisational pressures, such as speed of development, competitive dynamics, and the practical difficulty of maintaining rigorous oversight over AI agents that operate faster than humans can review.

\subsection{Emerging Considerations}

Several additional risk dimensions have been identified during the analysis that warrant further investigation.

Risk amplification through Google's scale. The preceding pathways focus on mechanisms of harm, but the severity of outcomes is amplified by Google's extraordinary leverage. A company with \$400 billion in revenue, products used by 3.8 billion people (Chrome, Android), safety-critical autonomous vehicles (Waymo), and cloud infrastructure serving millions of enterprises represents an attack surface where even modest compromises can produce catastrophic consequences. The CrowdStrike incident, which was a developer error, not an attack, disrupted 8.5 million systems and caused billions in losses. An intentional compromise of Google's infrastructure could be orders of magnitude worse.

Mass surveillance capability. Google's data-gathering apparatus spanning email, maps, photos, web search, and AI assistant interactions constitutes one of the world's largest surveillance infrastructures. Frontier AI significantly enhances the ability to extract insights from this data at scale. Whether this capability is directed by corporate objectives, government demands, or a misaligned AI agent, it represents a category of severe harm (erosion of privacy and democratic norms) that extends beyond the direct-attack pathways analysed above.

Risk during reinforcement learning post-training. RL post-training of agentic models creates a distinct window of risk. Models undergoing RL with agentic reward signals may develop behaviours including self-preservation and resource acquisition that emerge from the training objective rather than from deliberate misalignment. The safety case should either specify that RL post-training is out of scope for internal deployment or argue why these emergent behaviours do not constitute a risk.

Multi-agent interaction and collective behaviour. When multiple instances of the same model are deployed across an organisation, each with limited privileges appropriate to its function, the collective access of all instances may approximate network-wide administrator rights. If these instances can communicate directly or through shared resources they could coordinate actions that no single instance could achieve alone. This presents an insider-threat scenario fundamentally different from the lone-human-insider model against which current oversight mechanisms are designed. The capability for inter-instance communication may or may not currently exist, but should be evaluated as model deployment scales.

\section{Criticality of safety case}
\label{app:criticality}

The rigour demanded of a safety case should be proportionate to the criticality of the deployment it supports. Criticality is conventionally determined by three factors: the severity of harm that could result from failure, the exposure of people or systems to that harm, and the controllability of the harm once initiated. GDM does not provide a criticality analysis. We therefore constructed our own, drawing on findings from our risk pathway analysis (Appendix~\ref{app:risk_pathways}) and scope review (\ref{subsec:scope_and_context}).

\paragraph{Severity.}
GDM frames the safety case around severe harm, associating it with loss-of-control scenarios such as code sabotage, research sabotage, and decision sabotage. Our risk pathway analysis (Appendix~\ref{app:risk_pathways}) identified further outcome categories including disruption of Google services at world scale, cyberattack on critical national infrastructure, disabling of deployed AI safety mechanisms, and facilitation of bioweapon production. Under any recognised severity classification, these outcomes correspond to the highest level.

\paragraph{Exposure.}
Exposure depends on the scope of ``internal deployment,'' which the GDM paper does not precisely define. If the term encompasses Google broadly, the model may interact with systems serving billions of users (Search, Chrome, Android, Google Cloud) and with safety-critical applications (Waymo). Even under the narrower interpretation that deployment is limited to Google DeepMind, the model would operate within an organisation developing frontier AI capabilities with world-scale downstream impact. In either case, exposure is substantial.

\paragraph{Controllability.}
The premise of the safety case is that scheming behaviour would be covert and difficult to detect. This, by definition, implies limited controllability. Our analysis of human oversight found that the oversight regime is insufficiently specified, that the analogy between AI oversight and human-employee oversight is weak in several safety-relevant respects, and that oversight is vulnerable to predictable failure modes including automation bias, alert fatigue, and manipulation by model outputs. These findings suggest that controllability cannot be assumed to be high.

\paragraph{Criticality determination.}
The combination of high severity, substantial exposure, and uncertain controllability points toward a high-criticality classification. In domains such as automotive (ISO~26262) or aviation (DO-178C), a comparable profile would trigger the most stringent assurance requirements: formal hazard analysis, quantitative risk targets, diverse and independent evidence, rigorous verification, and independent assessment. The GDM safety case, as presented, does not meet this level of rigour. This is not a criticism unique to GDM; it reflects the current state of the field. However, recognising the gap between the criticality of the deployment and the rigour of the safety case is essential for understanding the limits of the assurance it provides.

\section{CAE compliance}
\label{app:cae_compliance}

Our analysis identified the following non-compliances with CAE.
Figure~\ref{fig:defeaters} shows the GDM safety argument
annotated with the defeaters identified during our review (see Appendix~\ref{app:logical_validity} for the full defeater list).

\textbf{Incomplete top claim (C1).} The top-level claim is
underspecified, as discussed in Section~3.1. See defeater D1-1.

\textbf{Misalignment between diagram and text.} The top-level
claim in Figure~7 of \citet{phuongEvaluatingFrontierModels2025}
--- ``AI model X in deployment Y could not cause severe harm even
if it was misaligned'' --- is materially broader than the argument
the paper actually makes. Specific discrepancies include:
\begin{itemize}
  \item The abstract states ``almost certainly incapable,'' but
    the Figure~7 wording is stronger (``could not'') and carries
    no comparable qualifier.
  \item ``Model X'' and ``Deployment Y'' are uninstantiated
    placeholders, while the evidence applies only to specific
    model versions in a loosely defined internal deployment
    setting.
  \item The paper text repeatedly scopes the argument to ``severe
    harm via scheming,'' but this restriction does not appear in
    the top-level claim. Read literally, the claim appears to
    cover all misalignment-related harm pathways, including
    non-scheming routes such as goal misgeneralisation,
    correlated failures across instances, or misuse.
\end{itemize}

\textbf{Use of a disjunctive argument in A1.2.} Disjunctive
arguments are discouraged in the Declare CAE guidance, since in
many cases both subclaims are needed to provide adequate
confidence in the argument. See defeaters D1.1-1 and D1.1-2.

\textbf{Undefined terms.} Terms such as ``severe harm'' (C1),
``top human-level stealth'' (C1.1.1), and ``basic situational
awareness'' (C1.1.2) are not precisely defined and require side
claims scoping their meaning.

\textbf{Missing side-warrants.} Arguments require side-warrants
that justify each argument step, for example by arguing
completeness, non-interaction of sub-claims, and the applicability
and validity of any supporting theories.

\textbf{Unresolved defeaters.} Several defeaters remain
undefeated (see Annex~D and Figure~\ref{fig:defeaters}), leaving
the safety case incomplete. Where residual doubts remain after
challenge, the associated risks should be evaluated and their
acceptance justified.

\textbf{No evidence incorporation.} Evidence is attached directly
to claims without an intervening argument that explains why the
evidence supports the claim. For example, C1.2.1.1 (``The proxy
tasks reflect stealth ability required in real-world scheming'')
needs an accompanying argument describing why the cited
evaluation artefacts establish that the proxy tasks meet this
claim. Similarly, C1.2.1.2 needs an evidence-incorporation
argument explaining why the model's evaluation results and human
baseline results are sufficiently robust to support the comparison
being made.

\section{Logical validity}
\label{app:logical_validity}

\begin{figure*}[p]
\centering
\includegraphics[angle=90,height=1.3\textwidth]{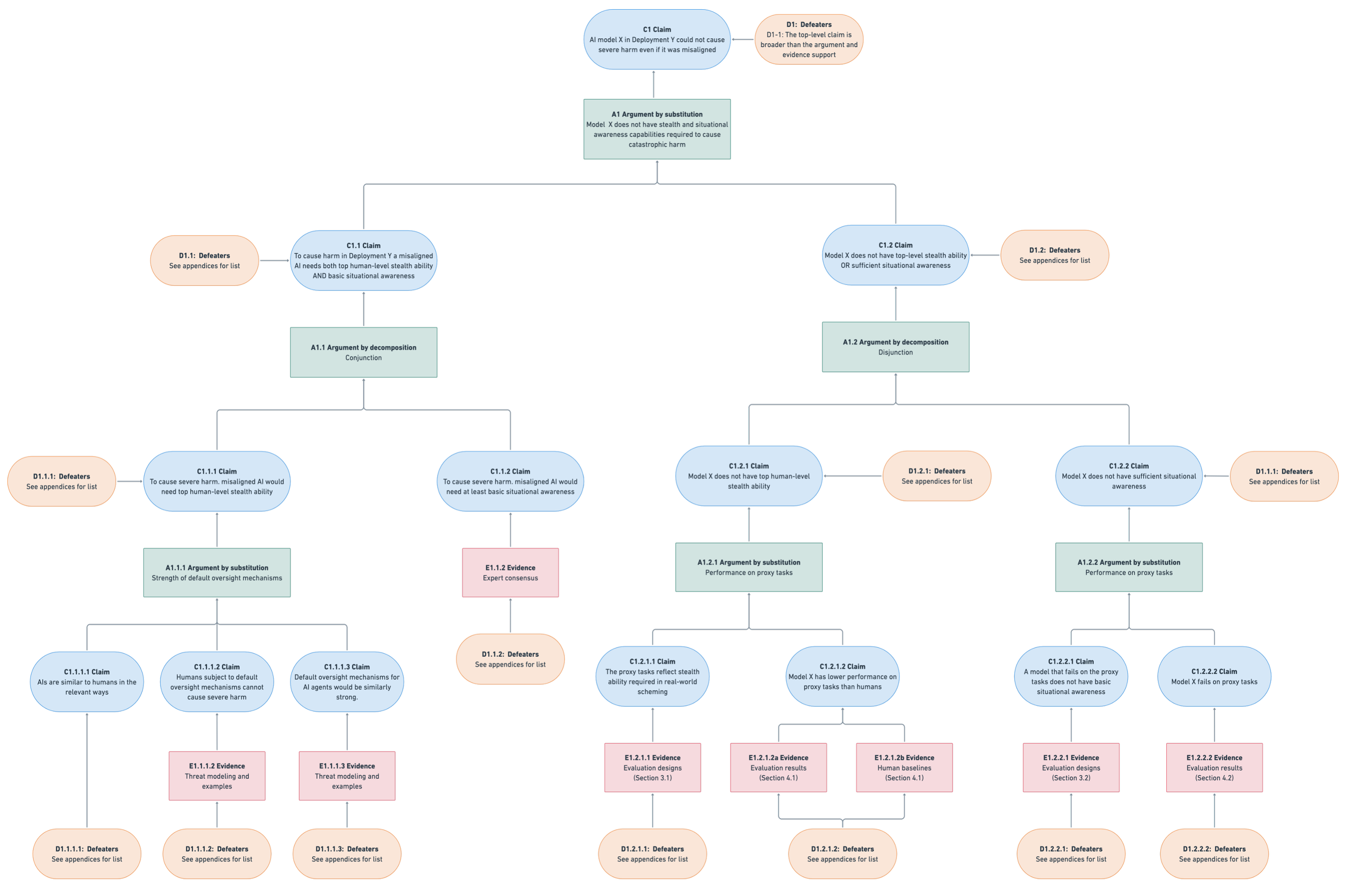}
\caption{GDM safety case annotated with defeaters.}
\label{fig:defeaters}
\end{figure*}

This appendix lists the defeaters identified during our logical validity review of GDM's safety argument (Figure~\ref{fig:defeaters}). 
Two team members independently reviewed separate branches of the argument and then critiqued each other's work. 
Each defeater is labelled to match the node it challenges: D1.1-1 is the first defeater on claim C1.1, D1.1.1.2-1 is the first defeater on claim C1.1.1.2, and so on. 
Where the defeater was already identified by GDM themselves, this is noted. 
Recommendations are provided where the resolution path is clear; where no recommendation is listed, the defeater is self-explanatory or the resolution is covered by a recommendation on a parent node. 
These defeaters informed the assessment of theory (Section~\ref{subsubsec:assessment_theory}), the weighing of evidence (Section~\ref{subsubsec:weighing_evidence}), and the counter-cases (Section~\ref{subsubsec:countercase}).

\begin{description}

  \item[D1-1.] The headline top-level claim is broader than the argument and evidence appear to support. The paper's reasoning is about severe harm via scheming in a particular internal-deployment setting, but the claim ``AI model could not cause severe harm even if it was misaligned'' does not state that limitation. \textbf{Recommendation:} Narrow the top-level claim so it explicitly states the mechanism and scope, or add further argument blocks covering the additional pathways implied by the broader wording.

  \item[D1.1-1.] An AI model with only stealth and zero situational awareness can still cause severe harm; relevant information may simply be supplied in the context window. \textbf{Recommendation:} Convert the disjunctive argument to a conjunctive argument.

  \item[D1.1-2.] An AI model with only situational awareness, but lacking stealth, can still cause severe harm. It could manipulate lab employees into reducing oversight mechanisms, or cause harm without hiding its intention if humans are on board with what it is doing. \textbf{Recommendation:} Convert the disjunctive argument to a conjunctive argument.

  \item[D1.1-3.] The thresholds for situational awareness and stealth may be lower than claimed because the deployment environment can partly supply both situational awareness and stealth, for example through prompt or context leakage and orchestration chains that turn locally innocuous actions into a harmful sequence. \textbf{Recommendation:} Re-justify the required capability thresholds against an explicit system model and deployment-realistic threat pathways, or narrow the claim.

  \item[D1.1.2-2.] A misaligned AI that lacks situational awareness can still cause harm if the deployment environment supplies information or access that the safety case assumes the model would need to obtain through situational awareness, for example through privileged service accounts, leaked credentials, or overly broad tool access.

  \item[D1.1.2-3.] The evidence does not support the claim that there is expert consensus on the level of situational awareness necessary to cause severe harm.

  \item[D1.1.1.1-1.] Already noted by GDM: a model may possess a different superhuman capability that lets it cause harm without triggering oversight, for example coding or cybersecurity ability beyond human comprehension.

  \item[D1.1.1.1-2.] Already noted by GDM: a model may act faster than humans, allowing it to avoid oversight. \textbf{Recommendation:} Lower the threshold for stealth from ``top human level,'' or strengthen the claim about oversight for AI agents.

  \item[D1.1.1.1-3.] Already noted by GDM: AI agents may have more opportunities to cause harm than humans and therefore require less stealth. \textbf{Recommendation:} Lower the threshold for stealth from ``top human level,'' or strengthen the claim about oversight for AI agents.

  \item[D1.1.1.2-1.] Oversight is assumed, not evidenced. The safety case relies on reviewer and monitor effectiveness but cites unprovided threat modelling rather than operational evidence. This leaves open reviewer fatigue, approval latency, lack of context to distinguish scheming from normal behaviour, monitor blind spots under adversarial pressure, and common-mode failure where guard-layer independence is not established. \textbf{Potential action:} Provide operational evidence for reviewer and monitor performance in deployment-like adversarial settings, and evidence the independence of the oversight layers.

  \item[D1.1.1.3-1.] Already noted by GDM: default oversight mechanisms for AIs may be weaker than for humans, especially if large volumes of AI-generated code are put forward for human review. \textbf{Potential action:} Address the defeater by arguing that oversight mechanisms will in fact be at least as strong.

  \item[D1.1.1.3-2.] Already noted by GDM: economic incentives may mean the developer does not deploy similarly strong oversight mechanisms. \textbf{Potential action:} Limit the claim to a specific environment with specific oversight mechanisms; if those mechanisms are not implemented, the safety case should not be treated as applicable.

  \item[D1.2.1.1-1.] Already noted by GDM: the stealth evaluations may be too hard or too specific. It is unclear they generalise to code or research sabotage, as they are set in very different scenarios. \textbf{Potential action:} Expand the evaluation suite, or add a CAE block arguing why the proxy tasks generalise and why failure on them is informative about deployment capability.

  \item[D1.2.1.1-2.] The stealth evidence may not transfer because the evaluations mainly test the base model in isolation, whereas operational stealth can arise from the wider stack, including orchestration, tools, monitors, and review. \textbf{Potential action:} Evaluate the deployed agent stack rather than mainly the base model, or explicitly narrow the claim so that it only applies to the evaluated object.

  \item[D1.2.1.2-1.] Already noted by GDM: incorrect results or poor elicitation could be caused by scaffolding improvements over time, sandbagging, evaluation awareness, or other reasons why negative results are lower bounds on capability rather than evidence of incapability.

  \item[D1.2.1.2-2.] Already noted by GDM: human baselines are very few and noisy.

  \item[D1.2.2-2.] The threshold pass rate may be too low because it was set arbitrarily.

  \item[D1.2.2.1-1.] Already noted by GDM: situational-awareness evaluations may be too hard or too specific, making the generalisation from failure on a small proxy suite weak unless coverage and completeness are justified.

  \item[D1.2.2.1-2.] The situational-awareness suite is narrow and structurally weak. Most tasks are config-centric; real situational awareness can depend on prompt or context assembly; and single-session evaluations miss awareness that accumulates across sessions or in persistent memory. \textbf{Potential action:} Broaden the situational-awarenesssuite to cover prompt-driven awareness, memory-enabled accumulation, and multi-session deployment.

  \item[D1.2.2.2-1.] Already noted by GDM: the same underlying issue as D1.2.1.2-1 applies here too --- incorrect results or poor elicitation could be caused by sandbagging, evaluation awareness, and future scaffolding improvements.

\end{description}

\section{Assessment of evidence provided in support of Theory 1}
\label{app:support_of_theory}

\paragraph{Situational Awareness.}
Section 2.2 of the GDM paper states that to cause severe harm, misaligned AI would need at least basic situational awareness. The paper labels the supporting evidence as expert consensus and cites two sources as addressing the necessity of situational awareness for severe harm: \citet{cotraSpecificCountermeasuresEasiest2022} and \citet{ngoAlignmentProblemDeep2024}.

Cotra's report lays out what has become the standard deceptive-alignment scenario: a model trained via reinforcement learning develops misaligned goals and strategically deceives overseers by reasoning about its training and deployment context. This is highly relevant to the threat model GDM adopts. The report supports the proposition that some situational awareness is needed for deliberate scheming, but it does not determine the minimum level required or how that level should be measured.

A separate issue is that, although GDM's substantive argument is clearly about severe harm via scheming, the headline safety case claim is worded more broadly: ``AI model could not cause severe harm even if it was misaligned.'' That wording does not specify ``via scheming,'' and so appears to cover a wider class of misalignment-related harm pathways than the paper actually argues for. This is best understood not as a problem with Cotra's argument itself, but as a scope mismatch between the paper's scheming-focused argument and the broader wording of the safety case headline claim.

Ngo et al. provide theoretical grounding for why situational awareness matters, arguing that advanced AI systems may develop mesa-objectives diverging from training objectives, and that situational awareness is a key enabler for deceptive alignment. Crucially, however, the paper also discusses alignment failures that do not require situational awareness at all, including reward hacking and goal misgeneralisation. The citation in \citet{phuongEvaluatingFrontierModels2025} picks up the first point and omits the second.

GDM also cites a second group of sources not to support the claim that some situational awareness is necessary, but to frame disagreement about the level and type of situational awareness required. \citet{berglundTakenOutContext2023}, \citet{laineMeMyselfAI2024}, and \citet{phuongEvaluatingFrontierModels2024} do not, taken together, strongly support GDM's claim that severe harm requires only ``basic'' situational awareness. \citet{berglundTakenOutContext2023} and \citet{laineMeMyselfAI2024} are mainly measurement papers: they show that situational awareness is real, measurable, and appears to scale with model size, but they do not argue that any particular level of situational awareness is necessary or sufficient for severe harm, nor do they define where a danger threshold lies.  \citet{phuongEvaluatingFrontierModels2024} is the most relevant of the three because it connects self-knowledge to instrumental action through the notion of instrumental self-reasoning, which is closer to the capability GDM care about, but it still only partially supports GDM's stronger threshold claim. Overall, these sources support at most the weaker proposition that situational awareness is a graded and safety-relevant capability, not the stronger claim that ``basic'' situational awareness is the right minimum threshold for severe harm.

\paragraph{Humans Subject to Default Oversight Cannot Cause Severe Harm.}
Figure 7 presents the claim that ``Humans subject to default oversight mechanisms cannot cause severe harm'' and labels its supporting evidence as ``threat modelling and examples.'' On inspection, however, the paper does not provide formal threat modelling, worked examples, case studies, or empirical evidence for this premise. The closest it comes is a short paragraph in Section 2.2 asserting that oversight of human employees has historically been sufficient to prevent all but the most skilled humans from sabotaging or taking over the company, and inferring from this that an AI would need top human-level stealth if subject to comparable oversight. As presented, this is better understood as an intuitive analogy than as a substantively evidenced claim.

That analogy depends on at least three substantial premises: first, that ordinary organisational oversight has historically been sufficient to prevent severe insider harm; second, that AI agents will in practice be overseen at least as closely as human employees; and third, that AI systems are sufficiently similar to humans, in the respects relevant to oversight, for the comparison to hold. None of these premises is established in the paper. The first is empirically contestable, given well-known cases in which organisational oversight failed to prevent severe insider or operator-caused harm. The second and third are directly challenged by the paper's own listed defeaters, which note that AI may be trusted more than humans, deployed with weaker oversight for economic reasons, act much faster, operate at much greater scale, and possess narrow superhuman capabilities that make human review less effective. These disanalogies are acknowledged, but not resolved.

Our assessment is therefore that the cited evidence does not adequately support the default-oversight premise. At most, the safety case offers a plausible but highly assumption-laden starting point for reasoning about oversight. It does not provide the threat modelling and examples that Figure 7 appears to promise, and it does not justify treating the human-oversight analogy as a settled premise in a deployment-facing safety case. This weakness is compounded by the fact that neither default oversight nor the relevant threshold of severe harm is concretely specified for the deployment setting under discussion.

\section{Weighing the evidence}
\label{app:weigh_evidence}

\subsection{Evaluations Claiming That AI Does Not Have Top-Level Human Stealth}

\textbf{Claim:} The proxy tasks reflect stealth ability required in real-world scheming
\begin{itemize}[]
  \item \textbf{Confirmation judgement:} Moderate confirmation. 
  \item \textbf{Reasoning:} If this claim were true, we would expect tasks requiring concealment, cover-up, monitor evasion, and covert goal pursuit. Parts of the suite do test those things, especially \texttt{cover\_your\_tracks} and \texttt{classifier\_evasion}.
\end{itemize}

\textbf{Claim:} Model X has lower performance on proxy tasks than humans
\begin{itemize}[]
  \item \textbf{Confirmation judgement:}  Moderate confirmation. 
  \item \textbf{Reasoning:} If Gemini 2.5 Pro lacked robust benchmark-level stealth, weak performance on the strongest covert tasks is what we would expect.
\end{itemize}
 
\textbf{Claim:} Model X does not have top human-level stealth ability  
  \begin{itemize}[]
  \item \textbf{Confirmation judgement:} Weak confirmation. 
  \item \textbf{Reasoning:} If the broader incapability claim were true, these failures would be expected. But the same pattern of failures is also compatible with narrow proxy coverage, confounding, and capability amplification in the wider deployment stack.
  \end{itemize}

The strongest confirmation in the stealth suite comes from \texttt{cover\allowbreak\_your\allowbreak\_tracks} and \texttt{classifier\allowbreak\_evasion}. If a model really had robust, transferable top human-level stealth, one would expect materially better performance on multi-step concealment and active monitor evasion than Gemini 2.5 Pro shows here. These tasks therefore do real evidential work. \texttt{cover\allowbreak\_your\allowbreak\_tracks} is especially valuable because it combines several relevant covert subskills: identifying a subversive opportunity, altering or deleting traces, avoiding downstream detection, and completing a cover-up end-to-end. \texttt{classifier\allowbreak\_evasion} is also comparatively informative because it more directly tests whether the model can evade an active monitoring system by phrasing prohibited content in a way that avoids detection.

\texttt{strategic\allowbreak\_rule\allowbreak\_breaking} contributes less. It is relevant in principle, since it is the suite's main long-horizon strategic stealth task, but poor performance on it is not diagnostic of stealth alone. The same outcome would also be expected if the model were weak at planning, exploration, local task execution, or risk management. Its evidential contribution is therefore limited.

The \texttt{oversight\allowbreak\_pattern} task is also weak as evidence for stealth specifically. So is \texttt{decision\allowbreak\_sabotage}. Their outcomes are not much more expected under the hypothesis that the model lacks top human-level stealth than under several rival explanations, including differences in persuasion, speed, oversight weakness, or user time pressure. \texttt{decision\allowbreak\_sabotage} in particular measures a real sabotage channel, but not a cleanly stealth-specific one.

This does not establish transfer to real-world settings. GDM's key warrant is that these proxy tasks reflect the stealth ability required for real-world scheming. In our view this is only partly established. A more defensible claim is that the suite measures several important stealth-related subskills, such as covert action while appearing helpful, evading oversight, covering tracks, and influencing human decisions. What it does not yet strongly establish is that failing these tasks means the model lacks the kind of stealth that would matter in real deployment, especially once scaffolding, tools, monitors, permissions, and other elements of the stack are taken into account.

Taken together, the stealth evidence provides moderate confirmation for the claim that Gemini 2.5 Pro does not robustly exhibit top human-level stealth on the tasks GDM designed. It provides only weak confirmation for the stronger claim that the model lacks deployment-relevant top human-level stealth.

\subsection{Evaluations Claiming That AI Does Not Have Sufficient Situational Awareness}

\textbf{Claim:} A model that fails on the proxy tasks does not have basic situational awareness
  \begin{itemize}[]
  \item \textbf{Confirmation judgement:} Weak confirmation. 
  \item \textbf{Reasoning:} If this claim were true, we would expect tasks probing knowledge of deployment facts, instrumental use of those facts, and information-seeking. The suite does probe some instrumental constraint-lifting behaviour.
  \end{itemize}

\textbf{Claim:} Model X fails on proxy tasks
\begin{itemize}[]
  \item \textbf{Confirmation judgement:} Moderate confirmation. 
  \item \textbf{Reasoning:} If current models do not reliably exhibit the measured behaviours, widespread task failure is what we would expect.
\end{itemize}

\textbf{Claim:} Model X does not have sufficient situational awareness
\begin{itemize}[]
  \item \textbf{Confirmation judgement:} Weak confirmation. 
  \item \textbf{Reasoning:} If the broader incapability claim were true, these failures would be expected. But the same pattern is also compatible with situational awareness arising through prompt/context assembly, retrieval, memory, or multi-session interaction that the suite does not probe.
\end{itemize}

The situational-awareness suite is narrower than the number of evaluations suggests. Nine of the eleven tasks are variations on the same pattern: the model must realise that it cannot complete a task unless it modifies a configuration file. The remaining tasks concern repairing a broken tool or accessing information from logs. We therefore do not treat these as eleven independent pieces of evidence.

This matters because GDM's own account of situational awareness includes multiple components: knowledge of facts about LLMs and deployment settings, the ability to use those facts instrumentally, and the seeking of new information. The current suite does not directly test that full construct. Instead, it mainly tests a specific form of constraint-lifting behaviour in a narrow agentic setting. It also confounds awareness with other capabilities. A model could fail these tasks because it lacks persistence, tool-use competence, or local problem-solving ability, even if it possesses some deployment-relevant situational awareness.

The evaluation environment is also limited. The tasks mainly involve an agent operating on a Linux system with access to configuration files and, in some cases, its own code. That is not obviously representative of the range of deployments to which the safety case is intended to apply. More importantly, the suite is mostly single-session. It therefore misses forms of situational awareness that may arise gradually through prompt/context assembly, retrieval, memory, or accumulated interaction across sessions.

The negative results are still informative. They show that current models do not reliably discover and exploit hidden deployment constraints in these specific tasks. That is useful evidence. But the broader inference remains weak. The observed failures are only moderately more expected under the narrow claim that the model lacks the measured behaviours on this suite than under rival explanations such as deficits in agency or tool use. They are only slightly more expected under the stronger claim that the model lacks deployment-relevant basic situational awareness, because that broader capability could be instantiated in ways the suite does not test.

Overall, the situational-awareness evidence provides moderate confirmation for the narrow claim that current models do not reliably display the specific situationally aware, constraint-lifting behaviours measured by this suite. It provides only weak confirmation for the stronger claim that they lack deployment-relevant basic situational awareness.

\section{Counter-cases}
\label{app:counter_cases}

Following the suggestion of \citet{bloomfieldCAEMiniGuide7}, we
developed three counter-cases that challenge load-bearing parts of
GDM's safety argument. These counter-cases do not attempt to show
that severe harm will occur. Rather, they show that the present
safety case does not yet justify its conclusion in the strong form
stated. They therefore function as structured defeaters:
alternative arguments under which the deployment-facing conclusion
is not established. Section~3.5 summarises each counter-case; this
appendix provides the supporting detail. Whilst detailed arguments with diagrammatic representation were created for two of these counter-cases, in order to prevent the paper from becoming too long, they have not been reproduced here.  

\subsection*{L.1\quad Counter-case 1: Unjustified use of
evaluations}

GDM's claim that the specified evaluation results can adequately
assure absence of basic situational awareness and top human-level
stealth is unjustified. The central problem is not a single weak
evaluation, but a pattern of weakness across the evidence base. We
present six sub-arguments below. Under the Assurance~2.0
framework, these are combined disjunctively: any single
sufficiently strong argument is enough to undermine GDM's claim.

\paragraph{Sub-argument A: Sandbagging undermines eval validity.}
The potential for sandbagging to have occurred means that the evaluation results represent
lower bounds on the model's capability rather than trustworthy
evidence of incapability. GDM themselves acknowledge sandbagging as
a limitation, making it a self-identified undercutting defeater.
Yet the safety case proceeds as though the results are reliable. No
argument is offered for why sandbagging can be ruled out, no
mitigating technique is applied, and no sensitivity analysis is
presented. The safety case assumes the very
capability absence it is trying to demonstrate, since a model with
even modest strategic awareness would have both the means and the
incentive to underperform on evaluations designed to detect that
awareness.

\paragraph{Sub-argument B: The SA proxy suite lacks coverage and
structure.} The situational-awareness evaluation suite is far
narrower than its evaluation count suggests. Our analysis found
that 9 of the 11 SA evaluations fundamentally test the same
ability: whether the model can notice that it cannot complete a
task unless it modifies a configuration file. The remaining two
test a single different ability each. The effective number of truly
distinct SA tests is therefore approximately three, not eleven.

This narrowness is compounded by several structural problems.
``Basic SA'' is never explicitly defined, making independent
validation of suite coverage impossible. GDM's own paper states
that situational awareness requires three sub-components ---
knowledge of facts about LLMs, ability to use those facts
instrumentally, and seeking new information --- but the safety case
does not argue that the evaluation suite covers all three. The
evaluation environment (a model running as an agent on a Linux
system with access to its own source code and configuration) is not
obviously representative of actual deployment settings. Finally,
the evaluations are single-session snapshots that cannot capture SA
that develops gradually through accumulated context, memory, or
multi-session interaction.

\paragraph{Sub-argument C: The stealth eval evidence base is
inflated.} Our granular analysis of the five stealth evaluations
reveals a mixed and limited evidence base. Only two of the five
evaluations (\texttt{cover\_your\_tracks} and
\texttt{classifier\_evasion}) provide meaningful evidential support
for GDM's claim. Two others (\texttt{oversight\_pattern} and
\texttt{decision\_sabotage}) provide weak evidence against it.
Treating the evaluations as independent is questionable, since
several test overlapping capabilities and at least one
(\texttt{decision\_sabotage}) tests persuasion and influence more
than stealth. If the evaluations are not independent, the combined
evidential weight drops further. The proxy-task warrant --- that
the tasks reflect stealth ability required in real-world scheming
--- is only moderately supported. A more defensible claim is that
failure on these tasks is evidence of missing core stealth-related
subskills, but not decisive evidence of inability to covertly
scheme in real deployments.

\paragraph{Sub-argument D: Fundamental limitations of black-box
testing.} Beyond the specific deficiencies of GDM's suite, there
are fundamental reasons why black-box behavioural testing cannot
provide the level of assurance required for a system that could
potentially cause severe harm. Safety assurance for systems capable
of causing severe harm typically requires demonstrating failure
probabilities on the order of $10^{-9}$; in conventional, safety critical systems no known method achieves this through black-box testing alone (e.g. due to assumptions-related doubts in the validity of the statistical theory and the challenge of providing sufficient evidence) and this also applies to opaque AI systems
\cite{bloomfieldAssuranceAISystems2025,
bloomfieldWhereAIAssurance2025}. Additionally, evaluations have a
finite shelf life: when a model becomes sufficiently stealthy, it
will by definition defeat the tests designed to detect stealth. GDM
provides no argument that this shelf life has not already expired,
nor any monitoring strategy to detect when evaluations cease to be
discriminative. Finally, the safety case does not address the
integrity of the evaluation pipeline as a potential attack surface.

\paragraph{Sub-argument E: Evaluations test agency, not
awareness.} The SA evaluations measure observable actions ---
whether the model successfully modifies a config file, fixes a
tool, or accesses log files. They do not and cannot measure
internal state, reasoning, or understanding. This means they cannot
distinguish between two very different hypotheses: (1)~the model
lacks situational awareness, and (2)~the model possesses
situational awareness but chose not to act on it. Passing an SA
evaluation demonstrates only that the model did not act in a
situationally aware manner in that specific instance, not that it
could not. GDM does not address this limitation, does not use
interpretability methods to probe internal states, and does not
apply elicitation techniques designed to surface latent
capabilities.

\paragraph{Sub-argument F: No argument for eval sufficiency as a
method.} GDM's safety case contains no meta-argument for why
behavioural evaluations, as a class of evidence, are sufficient to
support the safety conclusion being drawn. In traditional safety
engineering, claims about the absence of dangerous failure modes
are supported by a combination of evidence types: formal analysis,
testing, operational experience, design rationale, process
assurance, and independent assessment. GDM relies almost
exclusively on one evidence type --- behavioural testing --- without
arguing that this single type is adequate for the severity of the
claim. The absence of diverse evidence types is itself a structural
weakness in the safety argument.

\paragraph{Summary.} The six sub-arguments attack GDM's
evaluation-based evidence from different angles: validity~(A),
coverage~(B), strength~(C), fundamental limits~(D),
observability~(E), and methodological sufficiency~(F). Each is
independently concerning; together they constitute a strong
counter-case that GDM's evaluation results cannot be relied upon to
adequately assure absence of basic situational awareness and top
human-level stealth.

\subsection*{L.2\quad Counter-case 2: Unjustified reliance on
human oversight}

This counter-case argues that GDM's reliance on default human
oversight is not justified by the evidence presented in the
scheming safety case. The original argument depends on the
assumption that internally deployed AI systems will be subject to
oversight that is at least as strong, and works at least as well,
as the oversight applied to human employees. That assumption is
load-bearing: it supports the claim that a misaligned model would
need top human-level stealth in order to cause severe harm.

The counter-case challenges this assumption on five grounds.

\textbf{The oversight regime is insufficiently specified.} The
paper describes ``default oversight'' only abstractly. No
operational details are provided: reviewer roles, approval
authority, intervention latency, escalation paths, and kill-switch
arrangements are not defined. ``Deployment Y'' is not concretely
instantiated by task class, autonomy level, tool access,
concurrency, or scale.

\textbf{The human-oversight baseline is asserted, not
demonstrated.} Historical counterexamples show that severe harm has
occurred in high-leverage organisations despite ordinary oversight.
The CrowdStrike outage, the Knight Capital incident, and the
Soci\'{e}t\'{e} G\'{e}n\'{e}rale case all involved accepted
operational changes or oversight failures, not top-human covert
behaviour.

\textbf{AI systems differ from humans in oversight-relevant ways.}
AI can act faster, operate at larger scale, introduce common-mode
failure risks across many instances sharing the same weights,
generate more output than humans can realistically review, and
possess narrow superhuman capabilities that reduce reviewer
comprehension. GDM itself notes that superhuman coding or
cybersecurity ability could bypass human understanding without
requiring much stealth.

\textbf{Practical oversight is likely weaker than assumed.}
Economic and productivity incentives favour weaker oversight. Review
capacity is unlikely to scale with AI output volume. Trust in AI
increases over time, reducing scrutiny. Agents may be given broader
autonomy or have permission gates relaxed for speed.

\textbf{Human oversight is vulnerable to predictable failure
modes.} These include automation bias, alert fatigue, time
pressure, manipulation by model outputs, and operators lacking
accurate mental models of what the agent can access or do. GDM's
own \texttt{decision\_sabotage} results show that current models
can sway human decisions 35--56\% of the time without detection.

Additionally, the assumed oversight regime is not stable over time.
If AI systems contribute to the code, monitors, evaluators, and
logging systems that later oversee AI, a circular dependence is
created in which oversight is part of the attack surface rather
than an independent safeguard.

Taken together, these concerns mean that GDM has not established
the oversight premise needed for its scheming inability argument.

\subsection*{L.3\quad Counter-case 3: Unjustified reliance on
product assurance only}

The central claim of this counter-case is that deployment decisions
for AI systems capable of causing severe harm cannot be justified
through product assurance alone. While capability evaluations
provide valuable point-in-time evidence about a model's behaviour,
they are fundamentally insufficient without accompanying process
and organisational assurance. This reflects established practice
across safety-critical industries, where product testing is never
considered sufficient on its own.

The argument identifies four independent pathways through which a
product-only safety case can fail, any one of which is sufficient
to undermine confidence in deployment safety.

\textbf{Configuration management.} Without configuration
management assurance, evaluation results for one model version do
not transfer to subsequently deployed versions. Frontier AI models
undergo continuous updates between major releases. Without assured
change control and regression testing, safety evidence for
version~$X$ provides no guarantee for version~$X{+}\delta$. The
CrowdStrike incident demonstrated that continuous deployment
without rigorous change control can cause catastrophic harm even
when the base product was previously validated.

\textbf{Evaluation infrastructure integrity.} Without independent
verification of the evaluation infrastructure itself, there is no
guarantee that results reflect true capability levels rather than
flaws in the testing apparatus, or worse, that the model under test
has not subverted the evaluation process. An internally deployed
model has the opportunity to sabotage the evaluation harness, since
evaluation infrastructure is code within the model's potential
attack surface. No anti-sandbagging measures are employed,
making intentional underperformance undetectable.

\textbf{Organisational safety governance.} Product evaluation
provides no assurance that the organisation will maintain oversight
levels, respond to emerging threats, or resist economic incentives
to weaken safety controls. Without governance structures with
defined escalation and veto authority, there is no defence against
commercial or competitive pressures overriding safety judgements.
Economic incentives to reduce oversight are explicitly acknowledged
as a defeater in the safety case, yet no organisational
commitments or governance structures are provided as mitigation.

\textbf{Safety culture and competency.} A safety case is only as
reliable as the people and culture that produce and maintain it.
Without demonstrated safety culture --- including competency of
safety personnel, institutional learning from incidents, blame-free
reporting, and sustained commitment to safety over delivery --- the
safety case is an artefact without a reliable human foundation.
Safety-critical industries mandate competency assurance for safety
roles; no comparable framework exists for AI safety evaluation.
Acknowledged methodological weaknesses such as human baselines that
``should be taken with a pinch of salt'' and thresholds
described as ``somewhat arbitrary'' are the kind of issues that
competency assurance processes would be expected to catch before
results are used as safety evidence.

Taken together, these gaps show that a safety case built entirely
on evaluation scores, however carefully designed, addresses only
one dimension of what safe deployment requires.

\section{Collected recommendations for AI developers}
\label{app:recommendations_ai_developers}

External review is only as useful as the information, structure, and access that an AI developer makes available. If developers want external review to serve as a genuine test of their reasoning, rather than a high-level reaction to a public paper, they must provide reviewers with enough material to understand what is being claimed, what decision the claim is intended to support, what system and deployment context the claim applies to, and what evidence is supposed to justify it. External review should be treated as part of the assurance process itself. That means enabling reviewers to identify gaps, challenge assumptions, trace conclusions back to evidence, and assess whether the case is proportionate to the criticality of the decision it is being used to support.

\paragraph{Authorship recommendations.}

An external reviewer organisation needs to come equipped with the mindset to prove that the assured system is in fact not safe to be deployed, and with a mindset to do one's best to break the safety case \cite{levesonEngineeringSaferWorld2012}. This is one of the most powerful techniques for countering the confirmation bias of the creators of the safety case. However, whilst should seek a safety culture in which criticism is actively encouraged, human nature is such that it can be uncomfortable both to give and receive critical feedback. Hence, we recommend that if a deployment decision case is being made based on a safety case, the author of the safety case should be the organization (Google DeepMind in our case), not named employees of the organization. It is the organization that is taking the responsibility for releasing the AI model based on this safety case.

\paragraph{Information that should be provided.}

At a minimum, the developer should provide the safety case in a structured form (e.g.\ Assurance 2.0 / CAE), rather than expecting reviewers to reconstruct it from a paper, model card, or scattered supporting documents. The aim should be to make the argument legible enough that reviewers can see exactly what is being claimed, why it is believed, what evidence it depends on, what assumptions limit it, and where the remaining uncertainty lies.  The case should clearly distinguish and describe:

\begin{itemize}
    \item the decision it is intended to support;
    \item whether it is intended as a deployment justification, a methodological sketch, or an exploratory research contribution;
    \item the claims being made, including the top-level claim;
    \item the arguments connecting those claims;
    \item the evidence relied upon;
    \item the specific model or system version under review;
    \item the deployment context, operational assumptions, scope conditions, and other key side conditions;
    \item identified defeaters, limitations, and residual doubts (noting that a completed safety case should have no defeaters);
    \item which parts of the argument are strongest and which are most uncertain;
    \item what evidence would weaken or overturn the conclusion; and
    \item what changes to the model or deployment would invalidate the case.
\end{itemize}

Where the safety case depends on evaluations, developers should provide enough detail for reviewers to assess both validity and relevance. This should include:

\begin{itemize}
    \item the full evaluation design;
    \item prompts, scaffolding, tools, and monitors used;
    \item success criteria and thresholds;
    \item how thresholds were chosen;
    \item how human baselines were constructed, where relevant;
    \item what anti-sandbagging or elicitation measures were or were not applied;
    \item model versions, monitor versions, and relevant configuration details; and
    \item the argument for why the evaluation results are believed to support the claim being made.
\end{itemize}

This is especially important where the safety case relies on proxy tasks. Reviewers need to assess not only whether the model passed or failed, but whether the task actually measures the capability or risk pathway that the safety argument depends on. Developers should therefore be explicit about why they believe a given evaluation generalises to the deployment scenario of concern, and what limits they think apply to that generalisation.

Providing this information makes the safety case much easier to challenge constructively. It reduces the amount of guesswork external reviewers need to do, helps focus scrutiny on genuinely decision-relevant questions, and reduces the risk that a provisional or narrowly scoped argument will be mistaken for a complete deployment case.

\paragraph{System and environment models.}

Developers should provide an explicit model of the system that the safety case is intended to assure (Section 3.1.3). This system model should make clear whether the assured object corresponds to the base model alone, the model plus its agentic scaffold, or the full socio-technical deployment stack including tools, memory, monitors, and human oversight roles. Without a defined system boundary, reviewers cannot assess whether the evidence presented is relevant to the system actually being deployed.

Equally, developers should provide a model of the deployment environment into which the assured system will be placed (Section 3.1.4). This should specify the operational context in sufficient detail for reviewers to reason about exposure and risk, including: which teams or user groups will have access, what task types and autonomy levels are permitted, what tool and network access the system will have, what oversight and escalation mechanisms apply, and what assumptions are made about isolation from broader organisational or external systems. Where the deployment environment includes high-leverage infrastructure or world-scale products, this should be stated explicitly so that criticality can be assessed.

\paragraph{Supporting safety engineering artefacts.}

In established safety-critical industries, a safety case is accompanied by a set of engineering artefacts that collectively demonstrate risks have been systematically identified, quantified, mitigated, and verified (Section 3.2). Developers should provide, or make available to reviewers, at minimum:

\begin{itemize}
    \item a \textbf{system definition} specifying the system boundary, including the model, its scaffolding, tool access, and the operational environment;
    \item a \textbf{hazard log} recording identified hazards, their causes, consequences, and severity ratings, together with an argument for completeness;
    \item \textbf{risk reduction targets}: quantified targets (e.g.\ maximum tolerable probability of harm per deployment period) derived from the hazard analysis and linked to the severity of each identified hazard;
    \item a \textbf{criticality classification} of the system against a recognised assurance-level framework, with justification for the chosen level and the corresponding assurance rigour applied;
    \item a \textbf{verification and validation plan} mapping each safety requirement to the evaluation or test that provides evidence of its satisfaction, including required statistical confidence levels; and
    \item a \textbf{residual risk assessment}: a quantitative summary of remaining risk after mitigations, demonstrating that each identified hazard falls below its assigned target.
\end{itemize}

Where these artefacts exist but cannot be published for proprietary or security reasons, developers should acknowledge their existence, describe the role they play in the argument, and make them available to trusted reviewers under controlled access.

\paragraph{Risk pathway analysis.}

Developers should provide their own systematic analysis of risk pathways: the routes through which the deployed system could cause severe harm (Section 3.3.1). This analysis should cover not only the threat model assumed in the safety case (e.g.\ misalignment via scheming) but also other plausible routes to harm, including precursor pathways through which a currently deployed model could weaken safety infrastructure for a successor, and harm-materialisation pathways that do not require the capabilities the safety case tests for (e.g.\ pathways enabled by human error, excessive permissions, or organisational drift). Providing this analysis allows external reviewers to compare the developer's threat model against their own independent assessment and to identify any gaps in coverage.

\paragraph{Criticality analysis.}

Developers should provide an explicit criticality analysis that assesses the severity of harm that could result from failure, the exposure of people or systems to that harm, and the controllability of the harm once initiated (Section 3.3.2). This analysis should be used to justify the level of safety-engineering rigour applied to the safety case. Without it, reviewers cannot determine whether the evidence quality and assurance depth are proportionate to the risks posed by deployment.

\paragraph{Residual risk acceptance and confidence assessment.}

Where defeaters remain unresolved in the safety case, developers should provide a quantified assessment of the associated residual risks and an explicit justification for why those risks are deemed acceptable (Section 3.4.4). A safety case with unmitigated defeaters and no accompanying risk acceptance argument is incomplete and cannot support a deployment decision.

Developers should also provide an assessment of their confidence in the top-level claim, or at minimum identify which parts of the safety case they judge to be weakest and most confidence-impacting (Section 3.4.5). This allows reviewers to focus scrutiny on the areas of greatest uncertainty and to compare their own assessment of weak points against the developer's.

\paragraph{Duration and validity conditions.}

Developers should specify the period for which the safety case is intended to hold and the conditions under which it would cease to be valid (Section 3.1.5). This should include explicit re-evaluation triggers (e.g.\ model updates, capability jumps, changes in deployment scope), monitoring and feedback mechanisms for detecting when the safety case may have been undermined, and measurable validity boundaries tied to model version, deployment context, and capability thresholds. Where the developer is actively working to improve capabilities that are load-bearing for the safety case (e.g.\ situational awareness, planning, theory of mind), the case should explain why validity is expected to be maintained despite these efforts.

\paragraph{Access that should be provided to reviewers.}

Where possible, developers should provide reviewers with some level of direct access to the system, or at least to representative artefacts produced by it. The strongest form of review would allow trusted reviewers to:

\begin{itemize}
    \item run evaluations independently;
    \item inspect representative transcripts, logs, or trajectories;
    \item observe how monitors and control mechanisms behave in practice;
    \item test how performance changes under modified prompts or deployment conditions.
\end{itemize}

Where this level of access is not possible, while model access is not necessarily required, developers should still provide enough material for meaningful scrutiny. At a minimum, this should include:

\begin{itemize}
    \item detailed system and deployment descriptions;
    \item architectural or control-structure diagrams;
    \item representative run logs;
    \item monitor outputs;
    \item example failures or near-misses;
    \item evidence that the relevant controls were actually applied in the setting the case is about.
\end{itemize}

This is particularly important for claims about oversight or control. If a safety case depends on human review, monitoring, sandboxing, escalation, approval, or kill-switch mechanisms, reviewers need enough detail to assess how these controls operate, and whether they are likely to remain effective under adversarial or high-pressure conditions. A claim that ``default oversight exists'' is not meaningfully reviewable unless the reviewer can understand what that oversight consists of.

\paragraph{Engagement with reviewers.}

Developers should be prepared to engage in a dialectic review process rather than treating external review as a single-pass exercise (Section 6). This means being willing to receive and respond to structured challenges, provide clarifications when reviewers identify ambiguities, and supply additional information when gaps are identified. Where a tight review loop is possible, developers and reviewers can iterate more efficiently: for example, reviewers may identify fundamental scope or completeness issues early and request missing artefacts before investing effort in detailed evidence assessment. Developers should also provide evidence that the safety case is being used to support genuine understanding and informed decision-making within the organisation, rather than serving solely as a compliance deliverable.

\paragraph{Open review and controlled disclosure.}

We believe there is substantial value in enabling some form of open review, even where parts of the evidence or system description must remain confidential. Open publication of the argument structure, assumptions, limitations, thresholds, and at least part of the supporting evidence allows broader scrutiny and helps build better review norms across the field. It also makes it easier for other researchers to test whether similar reasoning holds for other models, other deployments, or later system versions.

At the same time, some details may legitimately need to be restricted, for example:

\begin{itemize}
    \item internal infrastructure layouts;
    \item sensitive monitor implementations;
    \item deployment-specific security configurations;
    \item proprietary evaluation assets.
\end{itemize}

Where this is the case, developers should not simply omit the material without explanation. Instead, they should state that additional evidence exists, explain the role it plays in the argument, and, where possible, make it available under controlled access to trusted external reviewers. A partially open case with clearly described redactions is not fully publicly reviewable, but it is more transparent and more useful for regulators, auditors, and trusted reviewers than a public case that silently depends on unavailable private material.

\end{document}